\documentclass[
 aps,
 prb,
 reprint,
 amsmath,
 amssymb,
 superscriptaddress,
floatfix
]{revtex4-2}

\usepackage[utf8]{inputenc}
\usepackage[T1]{fontenc}
\usepackage[dvipsnames]{xcolor}
\usepackage{soul}
\usepackage{graphicx}

\usepackage{siunitx}
\DeclareSIUnit\angstrom{\protect \text {Å}}

\usepackage{hyperref}
 \hypersetup{
     colorlinks   = true,
     citecolor    = blue,
     linkcolor    = blue,
     urlcolor     = blue
}

\PassOptionsToPackage{version=4}{mhchem}
\usepackage{mhchem}

\newcommand{\dd}{\mathrm{d}}

\newcommand{\mOH}{\muH}

\newcommand{\TN}{T_\text{N}}
\newcommand{\TC}{T_\text{C}}

\newcommand{\cax}{\ensuremath{c} axis}

\newcommand{\muO}{\ensuremath{\mu_0}}
\newcommand{\muH}{\ensuremath{\mu_0H}}

\newcommand{\Hperpc}{\ensuremath{H \perp c}}
\newcommand{\Hparac}{\ensuremath{H \parallel c}}

\newcommand{\Mat}{{M_\text{at.}}}

\newcommand{\Euion}{\ce{Eu^{2+}}}

\newcommand{\Hac}{\ensuremath{H_\text{ac}}}

\newcommand{\mueff}{\ensuremath{\mu_\text{eff}}}
\newcommand{\NA}{N_\text{A}}
\newcommand{\kB}{k_\text{B}}
\newcommand{\Thp}{\theta_\text{p}}
\newcommand{\muB}{\mu_B}

\newcommand{\ThD}{\theta_\text{D}}
\newcommand{\ThEi}{\theta_{\text{E}_i}}

\newcommand{\Cmag}{C_{\text{mag}}}
\newcommand{\Cph}{C_{\text{ph}}}
\newcommand{\Cel}{C_{\text{el}}}
\newcommand{\Cp}{C_{p}}
\newcommand{\CD}{C_{\text{D}}}
\newcommand{\CE}{C_{\text{E}}}

\newcommand{\Smag}{S_{\text{mag}}}

\usepackage{lipsum}

\begin{document}
\title{Complex magnetic properties of \ce{EuAgAs} single crystals}

\author{Karolina Kowalczyk}
\email[Corresponding author: ]{klpodgor@agh.edu.pl}
\author{Kamila Komędera}
\author{Janusz Przewoźnik}
\author{Łukasz Gondek}
\author{Czesław Kapusta}
\author{Wojciech Tabiś}
\affiliation{AGH University of Krakow, Faculty of Physics and Applied Computer Science, al. A. Mickiewicza 30, 30-059 Krak\'ow, Poland}

\author{Michał Babij}
\author{Lan Maria Tran}
\email[Corresponding author: ]{l.m.tran@intibs.pl}
\affiliation{Institute of Low Temperature and Structure Research, Polish Academy of Sciences, ul. Okólna 2, 50-422 Wrocław, Poland}

\author{Damian Rybicki}
\email[Corresponding author: ]{ryba@agh.edu.pl}
\affiliation{AGH University of Krakow, Faculty of Physics and Applied Computer Science, al. A. Mickiewicza 30, 30-059 Krak\'ow, Poland}




\begin{abstract}
EuAgAs is an antiferromagnetic topological material exhibiting intriguing magnetic behavior. We investigate its structural, magnetic, and local electronic properties using X ray diffraction, M\"{o}ssbauer spectroscopy, dc magnetization, ac susceptibility, and heat capacity measurements. The results confirm antiferromagnetic ordering below $\TN$ and reveal pronounced magnetic anisotropy and several field induced metamagnetic transitions. We construct the magnetic phase diagram of \ce{EuAgAs}, identifying several distinct magnetic regions. 
The field and temperature dependent behavior observed suggests a noncollinear magnetic structure in the low field regime. The sequence of field induced transitions resembles that observed in centrosymmetric rare earth compounds hosting skyrmion phases, suggesting that competing magnetic interactions may play an important role in stabilizing the observed magnetic states. 
\end{abstract}

\maketitle

\section{Introduction}
Numerous compounds containing rare-earth elements exhibit a wide range of fascinating physical properties. One such compound is EuAgAs, an antiferromagnetic topological material with a hexagonal crystal structure described by the $P6_3/mmc$ space group (ZrBeSi-type).
This type of symmetry is relatively common among the europium 111 family of compounds, including EuCuAs, EuCuP, and EuAuAs. Compounds belonging to this family can exhibit unusual magnetic structures, such as the helical magnetic structure reported for EuCuAs~\cite{Soh2024}.

A study by~\citeauthor{Laha2021}~\cite{Laha2021} on \ce{EuAgAs} has shown that the compound exhibits antiferromagnetic order below $T_N = \SI{12}{K}$, followed, for small magnetic fields applied parallel to the $a$ axis, by a weak metamagnetic transition. Additionally, a large topological Hall effect and a field-induced Weyl-fermion state were observed. Furthermore, another study reported a pressure-induced structural phase transition from $P6_3/mmc$ to $Pnma$ at an applied pressure of approximately \SI{4}{GPa}, which stabilizes the ferromagnetic-like state~\cite{Zhang2023}.

EuAsAg remains a subject of intense interest. 
Very recently, two independent neutron-diffraction studies have been undertaken to determine its magnetic ground state~\cite{Gazzah2026,Soh2026}. Such experiments are highly challenging due to the large neutron-absorption cross-section of europium, which substantially reduces the available neutron flux and complicates the collection of high-quality diffraction data. The fact that two independent studies have nevertheless been performed highlights the considerable interest in the magnetic properties and underlying physics of EuAgAs.
In the study by~\citeauthor{Gazzah2026}~\cite{Gazzah2026}, 
from the single-crystal time-of-flight unpolarized neutron diffraction measurements, a colinear antiferromagnetic structure of \ce{EuAgAs} was determined,
with a propagation vector of $\mathbf{q} = (0,0,0.5)$, corresponding to a doubling of the magnetic unit cell along the \cax, and an in-plane $\upuparrows\downdownarrows$ spin arrangement. However, the DFT calculations showed that ferromagnetic and altermagnetic states have energies just above the AFM ground state. 
On the other hand, based on the polarized neutron diffraction study,~\citeauthor{Soh2026}~\cite{Soh2026} proposed  that \ce{EuAgAs} exhibits long-range helical (non-collinear) magnetic order characterized by two temperature dependent incommensurate propagation vectors $\mathbf{q}_1=(0,0,0.5+\delta_1)$ and $\mathbf{q}_2=(0,0,0.5+\delta_2)$, observed below $T_\text{N1}$ and $T_\text{N2}$, respectively. Different magnetic structures proposed by different groups  for the same Eu-based compound is not unusual, even if  seemingly the same preparation method is used. 
This is the case of, e.g., \ce{EuSn2As2}~\cite{Li2019, Pakhira2021}, \ce{EuZn2As2}~\cite{Bukowski2022, Blawat2022}, and \ce{EuZn2P2}~\cite{Berry2022, Krebber2023, Rybicki2024}.



To gain a deeper understanding of the properties of \ce{EuAgAs} and to clarify its magnetic behavior, we conducted a comprehensive study that included temperature dependent x-ray diffraction and M\"ossbauer spectroscopy measurements, as well as heat capacity, dc magnetization, and ac susceptibility measurements as functions of temperature and magnetic field. The combination of these complementary techniques allows us to investigate the structural, magnetic, and local electronic properties of \ce{EuAgAs} and provides additional insight into the reported differences in its magnetic behavior.

\section{Experimental}
Single crystals of \ce{EuAgAs} were synthesized using the flux method, as described in detail in our previous work~\cite{Podgorska2024}.  
Starting materials: europium (Onyxmet, dendritic, \SI{99.99}{\%}, in mineral oil), silver (Specpure spectrographically standarised), arsenic (Alfa Aesar, \SI{99.999}{\%}), and bisumth (KOCH-LIGHT, \SI{99.999}{\%}) pieces were weighed in the atomic ratio Eu:Ag:As:Bi = 1:1:1:9. 
The weighed elements were then placed into an alumina crucible, which was subsequently sealed under vacuum within a quartz ampule. 
The ampule was heated in a chamber furnace to \SI{1100}{\degreeCelsius} in \SI{15}{h}, and maintained at this temperature for \SI{24}{h}. 
Subsequently, the cooling process was started and continued for \SI{250}{h} until $\SI{700}{\degreeCelsius}$.
As a result of synthesis, black-plate single crystals were obtained. The crystals were separated from the flux by centrifugation, and after that the amalgam method was used to remove any remaining flux from the crystal surface. The quality of the samples was checked by scanning electron microscopy (SEM) using a FEI Nova Nano 230 scanning electron microscope with the EDX option.

Temperature dependent x-ray powder diffraction (XRD) measurements were carried out with Panalytical Empyrean diffractometer. 
During the measurements, the sample position was adjusted against thermal displacement of the sample stage. 
The resulting diffraction patterns were subsequently refined using the FullProf Rietveld package~\cite{Rodriguez1993}.

Heat capacity measurements were performed using the two-$\tau$ relaxation method with the Heat Capacity options of the Quantum Design Physical Property Measurement System (PPMS-9) in the temperature range 1.84-\SI{296}{K}. The sample for heat capacity measurements was affixed and thermally linked to the platform using Apiezon N grease. 
A background signal from both the addenda and the grease (versus temperature) was also recorded.

Magnetic measurements, ac magnetic susceptibility and dc magnetization, were performed using PPMS-9 equipped with the ACMS or VSM option. The investigated temperature range was 2-\SI{300}{K} and the applied external dc fields were up to $\mOH = \SI{9}{T}$. 
For the measurements in PPMS with the ACMS option, the sample was secured with plastic discs inside a plastic straw, allowing measurements with the magnetic field applied parallel and perpendicular to the \cax. For measurements of the ac susceptibility, the driving field $\Hac = \SI{1}{mT}$ (\SI{10}{Oe}) and frequencies of $f = \SI{111}{Hz}$, \SI{555}{Hz}, and \SI{1111}{Hz} were used.

$^{151}$Eu M\"ossbauer spectra were collected at \SI{300}{K}, \SI{200}{K}, \SI{100}{K}, \SI{50}{K}, \SI{15}{K}, \SI{10}{K}, and \SI{4.2}{K} in standard transmission geometry using a conventional constant acceleration spectrometer with the $\ce{{}^{151}Sm(SmF3)}$ source. The \SI{21.5}{keV} $\gamma$-rays were detected with a Kr-filled proportional counter. The absorber was prepared as a powder form of the sample mixed with a BN carrier. The absorber thickness amounted to $\SI{32.5}{mg/cm^2}$. The spectra were analyzed by means of a least squares fitting procedure. To determine absorption line positions and relative intensities, we used numerical diagonalization of the full hyperfine interactions Hamiltonian. The isomer shift $\delta$ is given relative to the $\ce{^{151}Sm(SmF3)}$ source at room temperature.

\section{Results and Discussion}

\subsection{Single crystal characterization}\label{sec:Characterization}

\begin{figure*}[!ht]
    \centering
    \includegraphics[width=\linewidth]{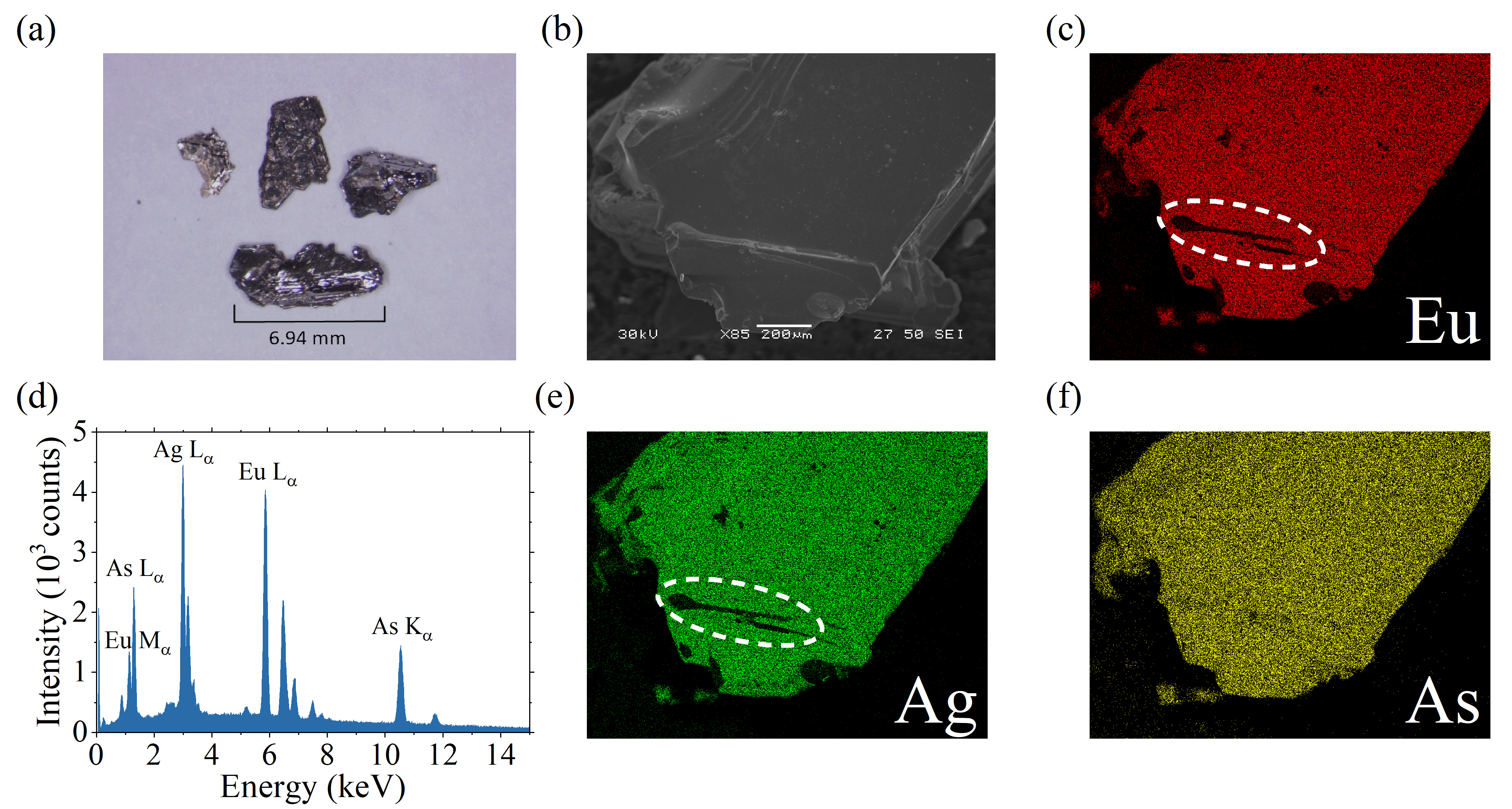}
    \caption{(a) Photo of example \ce{EuAgAs} single crystals, (b) SEM image, (d) EDX spectra, (c), (e) and (f) SEM images of \ce{EuAgAs} single crystal with individual elemental maps provided by x-ray microanalysis. Reproduced from: K. Podgórska et al., Synthesis of europium-based crystals containing As or P by a flux method: Attempts to grow EuAgP single crystals, {\it Solid State Sciences} 158 (2024) 107736, Copyright \textcopyright 2024 Elsevier Masson SAS. All rights reserved.}
    \label{fig:EDX1}
\end{figure*}

Examples of  crystals obtained after the surface cleaning procedure are presented in Fig.~\ref{fig:EDX1}(a).
They are characterized by a black plate-like appearance with surface sizes reaching several millimeters and thickness close to 0.1 millimeter.
The results of EDX analysis, shown in Fig.~\ref{fig:EDX1}d, provided the following elemental (atomic \%) composition: Eu 33.93, Ag 32.69, As 33.39. These values correspond very well to the expected stoichiometry. 
A SEM image of a selected crystal is shown in Fig.~\ref{fig:EDX1}b, while panels (c), (e), and (f) present elemental maps that confirm a uniform distribution of Eu, Ag, and As.
 Even after careful cleaning, small traces of Bi flux may remain on the surface. Due to the limited penetration depth of the electron beam, the measured signal is predominantly determined by the near surface region. Consequently, even small amounts of residual Bi flux can produce a noticeable Bi signal in the elemental maps and may obscure the determination of composition of the underlying material. This appears to be the case in our measurements, where the darker regions marked by the dashed lines in Fig.~\ref{fig:EDX1}(c) and (e) are likely associated with residual Bi flux.

\begin{figure*}[!ht]
    \centering
    \includegraphics[width=\linewidth]{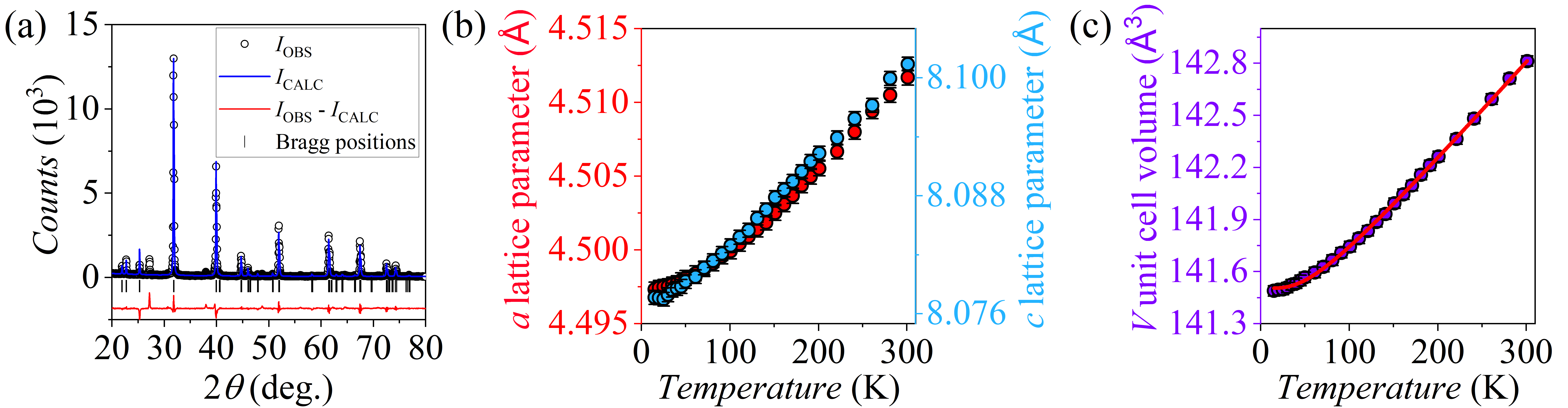}
    \caption{Results of x-ray diffraction studies in the temperature range 15-\SI{300}{K}. (a) Rietveld refinement (blue line) of diffraction pattern collected at \SI{300}{K} with Bragg positions marked by bars, reflection at $\SI{27}{\degree}$ of $2\theta$ originates from remains of Bi flux. (b) Temperature dependence of $a$ and $c$ lattice parameters, and (c) the unit cell volume $V$ with fitted Debye's curve, Eq.~\ref{eq:XRD_Debye}.}
    \label{fig:XRD1}
\end{figure*}

The results of XRD confirm that \ce{EuAgAs} crystallizes in the hexagonal $P 6_3/mmc$ space group (No. 194). The lattice parameters determined from the refined XRD pattern at $T = \SI{300}{K}$, Fig. \ref{fig:XRD1}(a), are $a = \SI{4.5122(1)}{\angstrom}$ and $c = \SI{8.1022(2)}{\angstrom}$, which agree well with the previous report~\cite{TomuschatSchuster1981}.  
Small non-indexed reflection around 27$^{\circ}$ of 2$\theta$ originates from the remaining Bi flux.

Due to limitations of the experimental setup, XRD measurements could not be performed below $\SI{15}{K}$. Consequently, it was not possible to investigate how the lattice parameters change around $\TN$. However, our results show that \ce{EuAgAs} does not undergo any structural phase transition between \SI{15}{K} and \SI{300}{K}. The lattice parameters $a$ and $c$ decrease monotonically with decreasing temperature.
Unit cell volume ($V$) was calculated from the lattice parameters and the resulting data were fitted using the Debye's formula: 
\begin{equation}
 V = V_0 + I_C \frac{T^4}{\theta ^3 _D} \int\limits_{0}^{\frac{\ThD}{T}} \dfrac{x^3}{e^x - 1} \,dx, 
 \label{eq:XRD_Debye}
\end{equation}
where $V_0$ is the unit cell volume extrapolated to 0 K, $I_C$ is a slope of the linear part of the $V(T)$ dependent on Gr{\"u}neisen and compressibility parameters, while $\ThD$ is the Debye temperature.
From the fit we obtained: $V_0 = (141.50 \pm 0.01)~\si{\angstrom}^3$, $I_C = (0.017 \pm 0.001)~\si{\angstrom^3/K}$ and $\ThD = (212 \pm 6)~\si{K}$.

\subsection{Heat capacity }\label{sec:HC}

\begin{figure*}[!ht]
    \centering
    \includegraphics[width=1\linewidth]{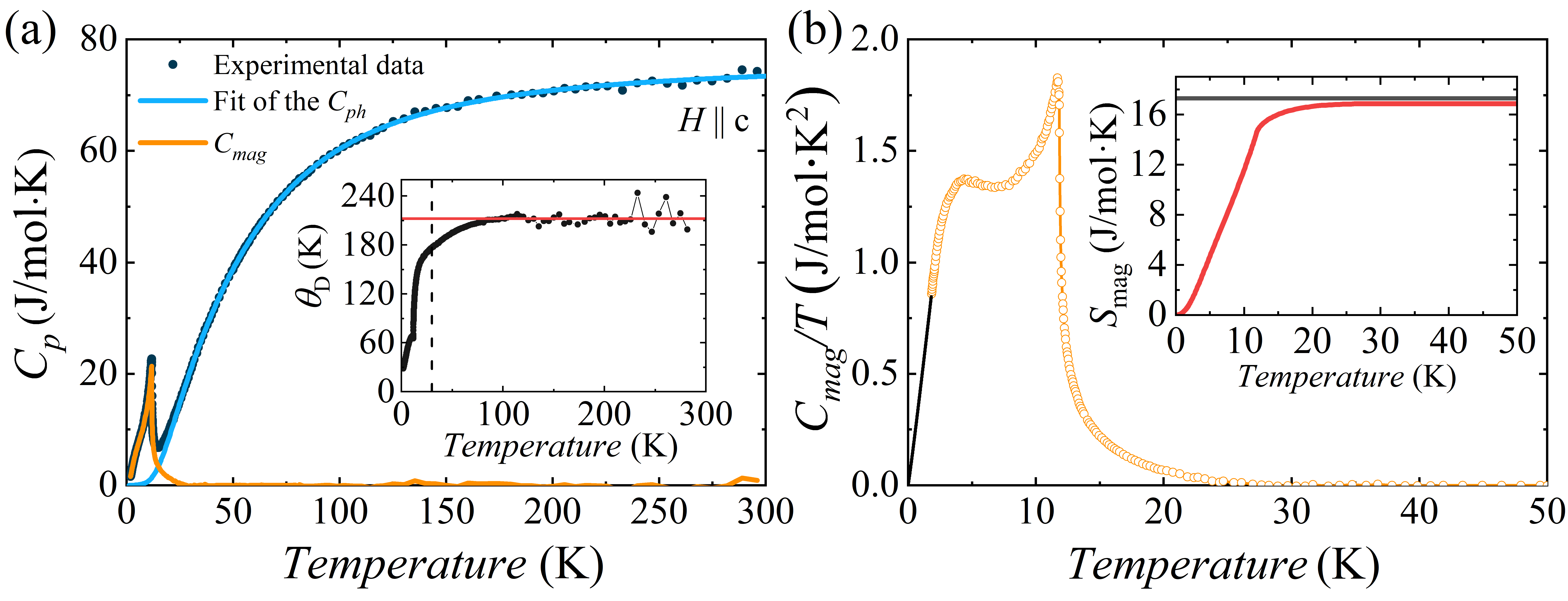}
    \caption{(a) Temperature dependence of specific heat $\Cp$ (solid points). The solid blue line illustrates the phonon contribution ($\Cph$), which was calculated using Eq.~\ref{eq:Cph}. The solid orange line represents the magnetic contribution ($\Cmag$) to the heat capacity determined as a difference between $\Cp$ and $\Cph$. The inset shows variation of the equivalent Debye temperatures $\ThD$ as a function of temperature. The dashed vertical line denotes the temperature above which magnetic contribution to $\Cp$  vanishes. A horizontal solid red line represents the fit to the experimental data points above $T=\SI{80}{K}$. Panel (b) shows the temperature dependence of $\Cmag/T$ (orange points) with extrapolation below \SI{1.84}{K} (black line) used for calculation of $\Smag$, which is shown in the inset. The black line in the inset corresponds to theoretical value of $\SI{17.28}{J.mol^{-1} K^{-1}}$.
    }
    \label{fig:Cp}
\end{figure*}
In general, the measured heat capacity can be expressed as a sum of several contributions:
\begin{equation}
\label{eq:Cp}
    \Cp = \Cph + \Cmag + \Cel
\end{equation}
where $\Cph$, $\Cmag$, $\Cel$, denote the phonon (lattice), magnetic, and electronic contributions, respectively. In the present analysis, the heat capacity was approximated by the sum of phonon and magnetic contributions, while the electronic contribution was neglected. 
Due to the presence of magnetic heat capacity the value of $\gamma$, in the electronic contribution $\Cel = \gamma T$, could not be reliably determined from the low-temperature experimental data, therefore $\gamma$ was fixed at zero in the final fit. 

The temperature dependence of the measured heat capacity $\Cp(T)$ and the $\Cph$ and $\Cmag$ contributions of \ce{EuAgAs} are shown in Fig.~\ref{fig:Cp}. 
A sharp peak appears at about \SI{11.7}{K}, which is the signature of long-range antiferromagnetic ordering of \Euion{} magnetic moments occurring at this temperature, which is in agreement with a previous report~\cite{Laha2021}. 
We assume that sufficiently far above this magnetic ordering the $\Cp$ vs $T$ dependence can be approximated by the phonon contribution using the following expression~\cite{Ashcroft, Gondek2007}, which contains the Debye's $\CD$ and Einstein's $\CE$ terms: 
\begin{equation}
\label{eq:Cph}
\Cph=\dfrac{R}{1-\alpha T}(\CD + \CE),
\end{equation}
where 
\begin{equation}
\CD = 9 \left( \dfrac{T}{\ThD} \right)^3 \int\limits_{0}^{\frac{\ThD}{T}} \dfrac{x^4 e^x}{ \left( e^x - 1 \right)^2} \,dx
\end{equation}
\begin{equation}
\CE = \sum_{i} \frac{m_i \left( \dfrac{\ThEi}{T} \right)^2 e^{\ThEi/T} } {\left( e^{\ThEi/T}-1 \right)^2},
\end{equation}
and $\ThD$ is a Debye temperature, $\ThEi$ are Einstein temperatures and $m_{i}$ are corresponding multiplicities for each individual optical branch, $\alpha$ stands for an anharmonic coefficient and $R$ is the universal gas constant.

In order to facilitate analysis, the summation over 6 independent optical branches was grouped into 2 branches with a 3-fold degeneracy. The fit to the experimental data was performed from 27 to 296 K, a range relatively far from the antiferromagnetic peak observed around 11.7 K, and is denoted by solid blue curve in Figure~\ref{fig:Cp}. 
The corresponding fit parameters are collected in Table~\ref{tab:Tab1}. 
The obtained Debye temperature of $151\pm \SI{9}{K}$ is in reasonable agreement with the value obtained from XRD measurements shown above ($212 \pm \SI{6}{K}$).
The lower Einstein temperature, $\theta_{\text{E}_1} = \SI{97}{K}$ might correspond to phonons of heavy Eu and the higher one, $\theta_{\text{E}_2} = \SI{257}{K}$ to lighter As atoms.

\begin{table}[!ht]
\caption{Debye ($\ThD$) and Einstein ($\ThEi$) temperatures and anharmonic coefficient ($\alpha$) obtained from the fitting of Eq.~\ref{eq:Cph} to the experimental data of the specific heat.}
\label{tab:Tab1}
\begin{tabular}{c| c} 
 \hline \hline
 $\ThD$ (K) & $151 \pm 9$  \\ \hline
 $\theta_{\text{E}_1}$ (K) & $97\pm 5$  \\ \hline
 $\theta_{\text{E}_2}$ (K)& $257\pm 3$  \\ \hline
 $\alpha$ (1/K)& $(2.5\pm 0.4)\times 10^{-5}$  \\ \hline \hline
\end{tabular}
\end{table}

In order to directly compare the Debye temperature obtained from heat capacity and the one obtained from the XRD measurements, we followed M.~Blackman's approximation ~\cite{Blackman1941}, where the heat capacity $\Cp(T)$ is given only by the Debye's term:
\begin{equation}
\label{eq:CpBlackman}
\Cp(T) = 9sR \left( \dfrac{T}{\ThD} \right)^3 \int\limits_{0}^{\frac{\ThD}{T}} \dfrac{x^4 e^x}{ \left( e^x - 1 \right)^2} \,dx,
\end{equation}
where $s$ is the number of atoms per molecule and the anharmonic coefficient $\alpha$ is additionally assumed to be zero.
The variation of the equivalent Debye temperature $\ThD$, calculated from Eq.~\ref{eq:CpBlackman}, as a function of $T$ is shown in the inset of Fig.\ref{fig:Cp}(a).  
One should also note that a perfect fit of the $\Cp(T)$ dependence with the Debye curve would imply a constant value of $\ThD$ as a function of $T$ (which is clearly not the case here).
However, for temperatures above approximately \SI{80}{K}, $\ThD$ appears to be independent of temperature. 
In this range, a horizontal straight line fit (solid red line in the inset of Fig.\ref{fig:Cp}(a)) yields $\ThD = 212.2\pm\SI{1.3}{K}$. 
This is in excellent agreement with the $\ThD=212\pm\SI{6}{K}$ value obtained from the XRD data analysis.

\begin{figure*}[!ht]
    \centering
    \includegraphics[width=1\linewidth]{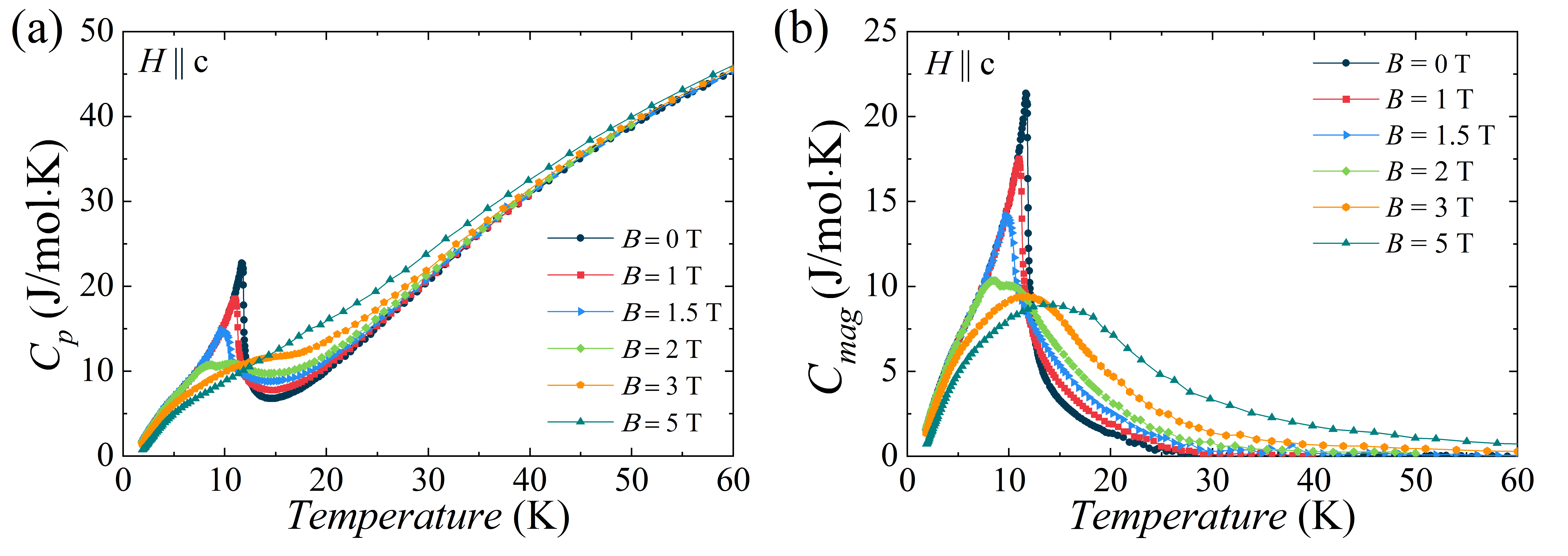}
    \caption{Temperature dependence of (a) specific heat capacity $\Cp$ and (b) its magnetic contribution $\Cmag$ measured at different magnetic fields applied along crystallographic \cax.}
    \label{fig:HC2}
\end{figure*}

The solid orange curve in Figure~\ref{fig:Cp} shows the magnetic part of specific heat, calculated using the formula: 
\begin{equation}
\Cmag(T) = \Cp(T) - \Cph(T).
\label{eq:Cmag}
\end{equation}
The intensity of the $\Cmag$ peak at $\TN$ amounts to $\SI{22.70}{J.mol^{-1} K^{-1}}$, which is close to the value of $\SI{20}{J.mol^{-1} K^{-1}}$ predicted by the mean field theory for an antiferromagnet of the equal magnetic moments value~\cite{Blanco1991}. 
From Fig.~\ref{fig:HC2}(b) one can notice that $\Cmag$ remains finite also above $\TN$ and vanishes only at temperatures roughly two times higher than $\TN$ (at about \SI{25}{K}), which is often explained as a signature of short range magnetic correlations present already above $\TN$~\cite{Pakhira2023, Rybicki2024, Podgorska2025}.

The magnetic contribution to the entropy, $\Smag(T)$, was derived from the $\Cmag(T)$ calculated from our experiment using the formula: 
\begin{equation} \label{eq:Sm}
\Smag(T)=\int\limits_{0}^{T} \dfrac{\Cmag}{T'} \,dT'. 
\end{equation}

Since we have the experimental data only above temperature \SI{1.84}{K}, the $\Cmag/T$ data were extrapolated from \SI{0}{K} to \SI{1.84}{K} using a power function in the form of $y=ax^b$ (fitted to the low-temperature experimental points) with $a= 0.45\pm 0.01$ and $b=1.06 \pm 0.03$. 
The temperature dependence of $\Smag(T)$ is shown in Fig.~\ref{fig:Cp}(b).
At $\TN$ it amounts to $\SI{14.34}{J.mol^{-1} K^{-1}}$ and with increasing temperature it reaches $\SI{16.84}{J.mol^{-1} K^{-1}}$. 
This is very close to the theoretical value of $\SI{17.28}{J.mol^{-1} K^{-1}}$, expected for one mole of particles with spin $J$ in a magnetic field given by $\Smag(T)=R\ln(2J+1)$, for $J = 7/2$ (value for \Euion{} ions). 
This indicates that Eu in \ce{EuAgAs} is present as \Euion{} ions.

The temperature dependence of $\Cp$ and $\Cmag$ in different magnetic fields up to \SI{5}{T} are shown in Fig.~\ref{fig:HC2}(a) and (b), respectively.
Due to the sample geometry, field dependent measurements could only be performed with the external magnetic field applied along the crystallographic \cax.

Fig.~\ref{fig:HC2}(b) presents the temperature and field dependence of $\Cmag$.
The sharp peak observed at $\TN$ in zero field shifts towards lower temperatures with increasing magnetic field, which is characteristic of antiferromagnetic order.
This shift is observed for magnetic fields up to \SI{1.5}{T}. 
At \SI{2}{T} the antiferromagnetic peak is still visible (at \SI{8.5}{K}), but another one appears at about \SI{10}{K}. 
At 3 T the antiferromagnetic peak is not visible, and the peak previously seen at \SI{10}{K} moves towards higher temperature and significantly broadens, which reminds behavior of a ferromagnet. 
This splitting of the zero Tesla peak is an indication of the reorientation of magnetic moments induced by the magnetic field~\cite{BEDNARCHUK2015, Podgorska2025}.

\subsection{$^{151}$Eu M\"ossbauer spectroscopy}\label{sec:MS}

\begin{figure}[!ht]
    \centering
    \includegraphics[width=1\linewidth]{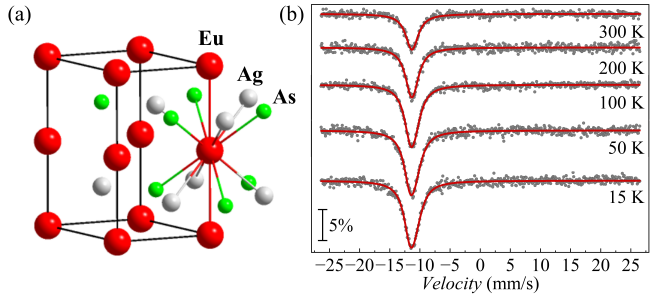}
    \caption{a) Unit cell and the local surrounding of Eu in EuAgAs, (b) $^{151}$Eu M\"ossbauer spectra measured at 300 K, 200 K, 100 K, 50 K, and 15 K, i.e., in the paramagnetic state, with their fits (solid lines).}
    \label{fig:Moss_H_T}
\end{figure}

To gain a deeper understanding of magnetic properties and local environment of Eu, especially information about the orientation of Eu magnetic moments, we conducted $^{151}$Eu Mössbauer spectroscopy measurements.
The results of the measurements above the transition temperature are presented in Fig.~\ref{fig:Moss_H_T}.
In the paramagnetic state in the presence of the electric quadrupole interaction for $^{151}$Eu there are 8 allowed transitions and the corresponding lines in the spectrum.
Their position is given by the following parameters: the isomer shift, $\delta$ and the main component of the electric field gradient tensor (EFG), $V_{ZZ}$.
The local surrounding of Eu is shown in Fig.~\ref{fig:Moss_H_T}(a) and Eu has a $-3m$ site symmetry, which indicates that the \cax{} is a threefold rotation axis and becomes the main component axis of the electric field gradient tensor ($V_{ZZ}$), which has an axial symmetry~\cite{Yoshida2013}.

\begin{figure}[!ht]
    \centering
    \includegraphics[width=1.0\linewidth]{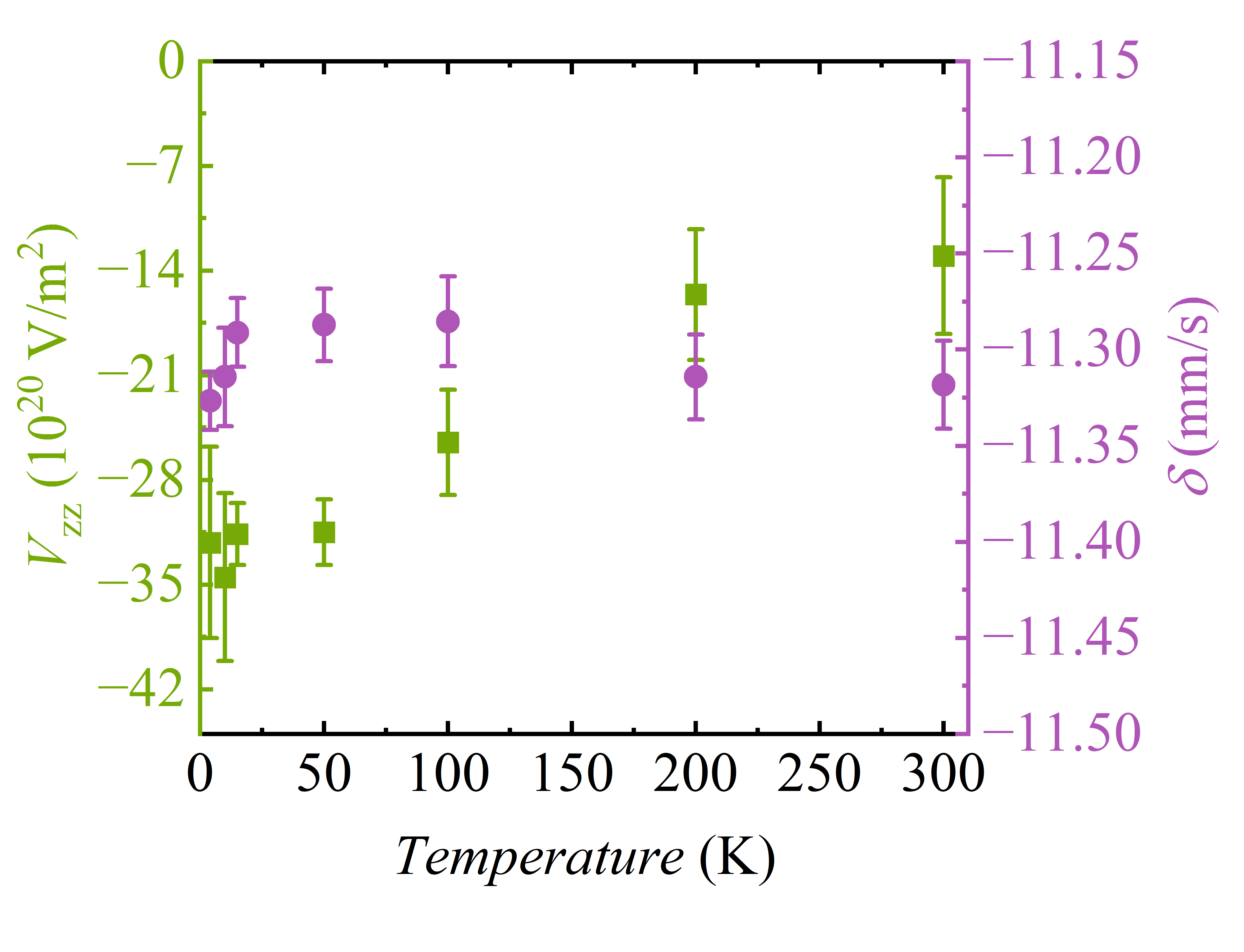}
    \caption{Temperature dependence of the main component of the electric field gradient $V_{ZZ}$ (green points) and the isomer shift $\delta$ (purple points).}
    \label{fig:Moss_Vzz_IS}
\end{figure}

From the room temperature spectrum we obtained $\delta=\SI{-11.32}{mm/s}$, and $V_{ZZ}=-13.01  \times 10^{20}~\si{V/m^{2}}$.
Values of $\delta$ around \SI{-10}{mm/s} are typical for europium in the Eu$^{2+}$ state while around \SI{0}{mm/s} for Eu$^{3+}$ state~\cite{Schellenberg2010, Komedera2021, Bukowski2022}.
No signal at approximately \SI{0}{mm/s} confirms that Eu is only in the Eu$^{2+}$ state and that there are no impurities containing Eu$^{3+}$.
Regarding $V_{ZZ}$, its sign cannot be determined in the paramagnetic state, particularly when the spectrum is symmetrical, as in our case.
Therefore, the negative sign of $V_{zz}$ was determined by the calculation of skewness, and we also note that the negative sign was concluded for EuCuAs and EuCuP compounds, which have the same crystal structure as EuAgAs~\cite{May2023, Ryan2025}.
The obtained value of $V_{ZZ}$ is small, which is due to the rather high symmetry of Eu site, compared to, e.g., EuSnP, where below and above Eu site there are different ions~\cite{Podgorska2025}.
Next we carried out measurements at lower temperatures (200, 100, 50, 15,  10 and 4.2 K). Before discussing these measurements in detail, it is worth noting that the isomer shift does not change significantly upon cooling, as shown in Fig.~\ref{fig:Moss_Vzz_IS}, suggesting that the electronic density at the Eu nuclei remains essentially unchanged.
The absolute value of $V_{ZZ}$ increases with decreasing temperature (Fig.\ref{fig:Moss_Vzz_IS}), as could be expected due to the lattice contraction and the corresponding decrease of the interatomic distances, as the XRD data presented above show in Sec~\ref{sec:Characterization}.


\begin{figure}[!ht]
    \centering
    \includegraphics[width=1.0\linewidth]{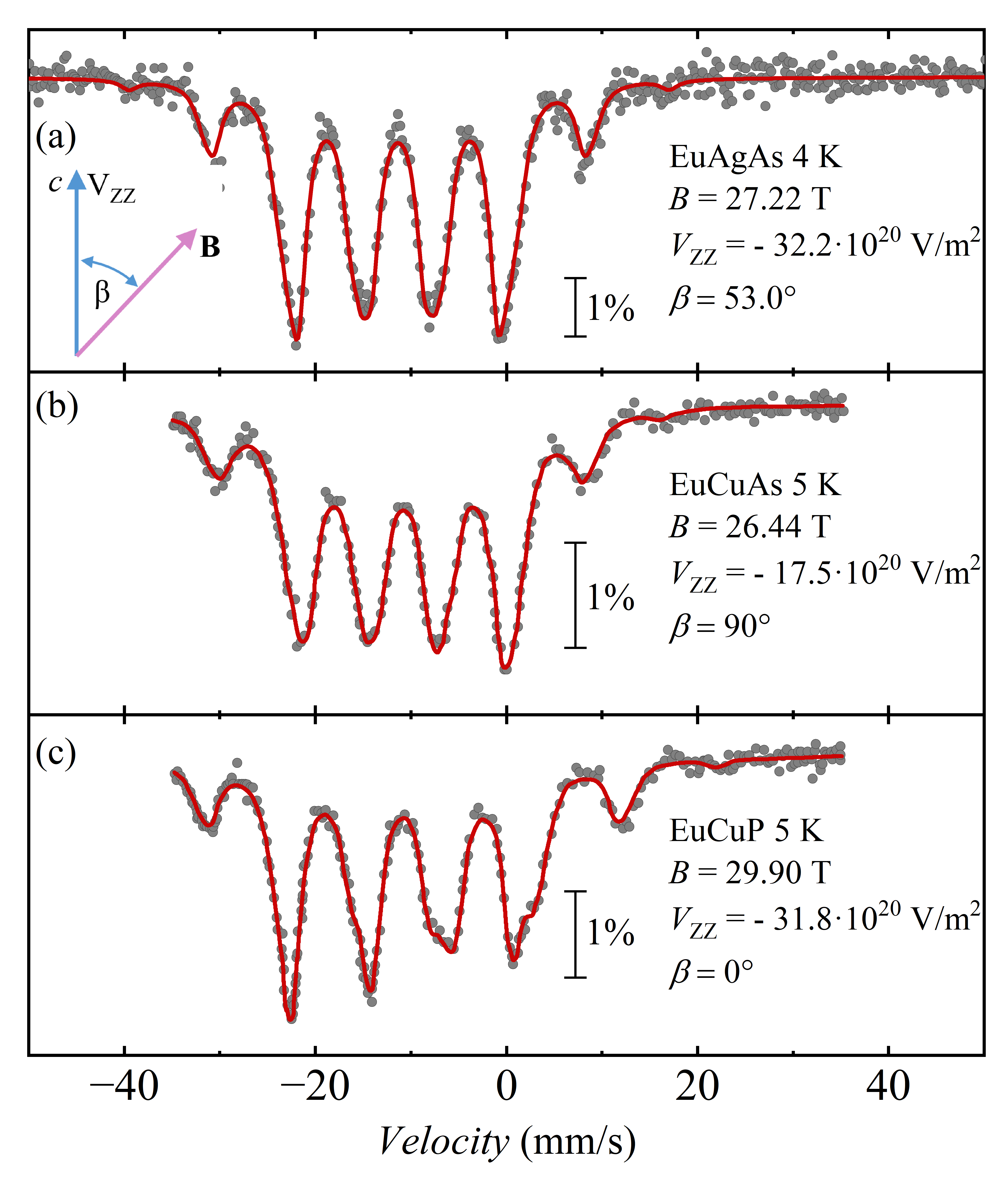}
    \caption{$^{151}$Eu M\"ossbauer spectra and fits (solid lines). Panel (a) shows result for \ce{EuAgAs} at 4 K. Panels (b) and (c) were adapted from D. H. Ryan (2025)~\cite{Ryan2025} and show the spectra for EuCuAs and EuCuP at 5 K. Values of hyperfine magnetic field $B$, the main component of electric field gradient tensor $V_{ZZ}$, and $\beta$ angle are included in each panel.}
    \label{fig:Moss_low_T}
\end{figure}

In the magnetically ordered state the magnetic field present at Eu nucleus causes further splitting of nuclear energy levels, which leads to more features observed in the spectrum. 
This can be seen in Fig.~\ref{fig:Moss_low_T}, where the measurement at 4 K is presented together with the literature spectra obtained for EuCuAs and EuCuP~\cite{Ryan2025}. 
From fitting such a spectrum one can additionally deduce the angle $\beta$ between the $V_{ZZ}$ axis and direction of the magnetic (hyperfine) field $B$, i.e., the direction of the Eu magnetic moment.
In our case, $\beta$ is the angle between the \cax{} and the direction of the magnetic moment.
Comparing the low temperature \ce{EuAgAs} spectrum with those of EuCuP and EuCuAs, it can be seen that the spectrum of \ce{EuAgAs} is the most symmetric one. 
For EuCuAs from neutron diffraction~\cite{Soh2024} it has been shown that its magnetic order is a planar helical with magnetic moments perpendicular to the \cax{} ($\beta=\SI{90}{\degree}$).
On the other hand, for EuCuP from M\"ossbauer spectroscopy it has been deduced that magnetic moments are parallel to the \cax{}~\cite{Ryan2025}, $\beta=\SI{0}{\degree}$.
For \ce{EuAgAs} we obtained the angle $\beta=(53\pm 1.4)\si{\degree}$.
Another parameter of the fit is the $V_{ZZ}=(-32.2\pm6.4) \times 10^{20}~\si{V/m^{2}}$, which is of the same order of magnitude as for EuCuP and EuCuAs.
The isomer shift obtained, $\delta=(-11.33 \pm 0.02)~\si{mm/s}$ is somewhat smaller (more negative) compared to the values reported for EuCuAs and EuCuP~\cite{May2023, Ryan2025}.
The hyperfine field value of $B= (27.22\pm 0.05)~\si{T}$ is higher compared to EuCuAs (\SI{26.44}{T}), but lower than that of EuCuP (\SI{29.9}{T})~\cite{Ryan2025}. 

\begin{figure}[!ht]
    \centering
    \includegraphics[width=1\linewidth]{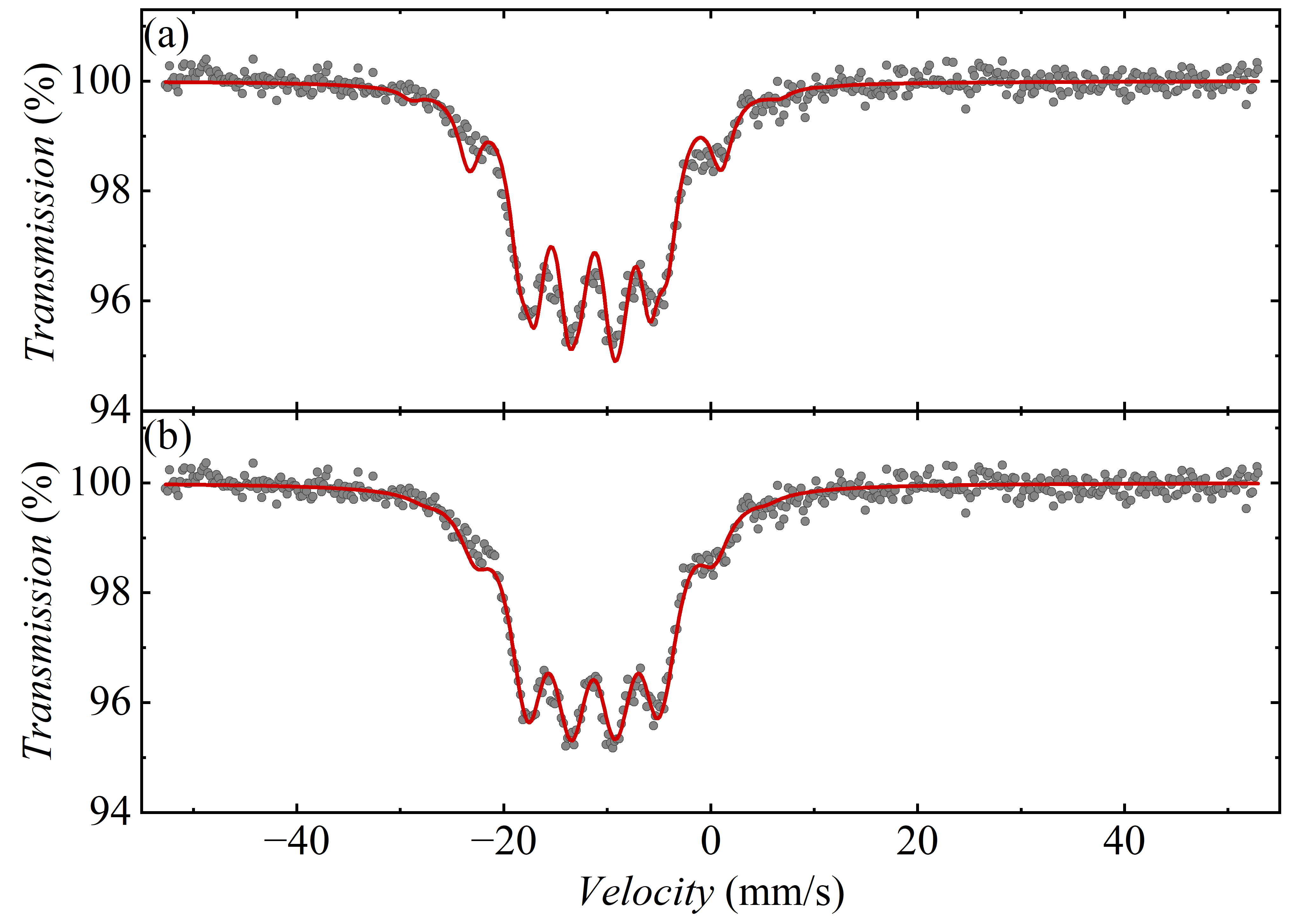}
    \caption{$^{151}$Eu M\"ossbauer spectra obtained at 10 K with their fits (solid lines). Panel (a) shows fit with smaller line width of 1.21 mm/s and (b) fit with larger line width of 1.71 mm/s.}
    \label{fig:Moss_10K}
\end{figure}

Before we proceed to discussion of the spectrum measured at \SI{10}{K}, i.e., just below $\TN$, we would like to note that the line width (half width at half maximum) was fitted for the spectrum at \SI{100}{K} and fixed at this value (\SI{1.21}{mm/s}) for other spectra in the paramagnetic state (\SI{300}{K}, \SI{200}{K}, \SI{50}{K}, and \SI{15}{K}).
Almost the same line width values were obtained for EuCuAs and EuCuP~\cite{May2023, Ryan2025}.
The spectrum measured at \SI{4}{K} was fitted using an even narrower line width of \SI{1.06}{mm/s}.
Initially, the spectrum at \SI{10}{K} was fitted using a fixed line width (\SI{1.21}{mm/s}) following the approach at higher temperatures, as shown in Fig.~\ref{fig:Moss_10K}(a).
However, the fit does not accurately describe the experimental data. 
We then allowed the line width to be a fitting parameter, which resulted in a better fit (Fig.~\ref{fig:Moss_10K}(b), but the fitted line width was significantly larger, equal to \SI{1.71}{mm/s}.
Fig.~\ref{fig:Moss_Vzz_IS} at \SI{10}{K} shows the values of $\delta$ and $V_{ZZ}$ obtained from a fit with a larger line width, as this approach provides better descriptions of the experimental data. 
In this fit, the angle $\beta$ remains the same, within the uncertainty, as at \SI{4}{K}.
In M\"ossbauer studies of EuCuP a similar effect of line broadening just below the Curie temperature ($\TC$) was observed and it was concluded that magnetic order in EuCuP likely passes through an incommensurate sinusoidally modulated state just below $\TC$~\cite{Ryan2025}. 

To conclude, our M\"ossbauer studies show that at very low temperatures Eu magnetic moments are tilted from the \cax{} by an angle of about 53$^{\circ}$ and we can speculate that a modulation of magnetic moments is present just below $\TN$.
We note that in the case of an axially symmetric EFG tensor, M\"ossbauer spectroscopy is not sensitive to  rotation of the magnetic moment about the \cax{}, i.e., such a change would result in a narrow line, therefore, a possible modulation would have to include a change of the angle $\beta$ or magnitude of the magnetic moments.

\subsection{Magnetic measurements}
We systematically investigated the magnetic properties of \ce{EuAgAs} by performing measurements of both ac susceptibility and dc magnetization as functions of temperature ($T$) at several external magnetic fields ($\muH$) and as a function of the external magnetic field at several temperatures. These measurements provide important insights into the nature of the magnetic order and its evolution with temperature and field, and are essential for constructing the magnetic phase diagram of EuAgAs.

The temperature dependence of the dc magnetic susceptibility, $\chi(T)$, was obtained from the dc magnetization data using 
\begin{equation}
\chi(T) = \dfrac{M(T)}{H},
\end{equation}
where $M$ is the magnetization and $H$ is the applied magnetic field. 
For the analysis of the field dependent measurements, the first (differential susceptibility, $\dd M/\dd H$) and second ($\dd ^2M/\dd H^2$) derivatives of the  magnetization with respect to the applied magnetic field were calculated.
Significant difference between the influence of the magnetic field applied parallel and perpendicular to the \cax~ is observed.

\subsubsection{Temperature dependence}
\begin{figure*}[!ht]
    \centering
    \includegraphics[width = 0.49\textwidth]{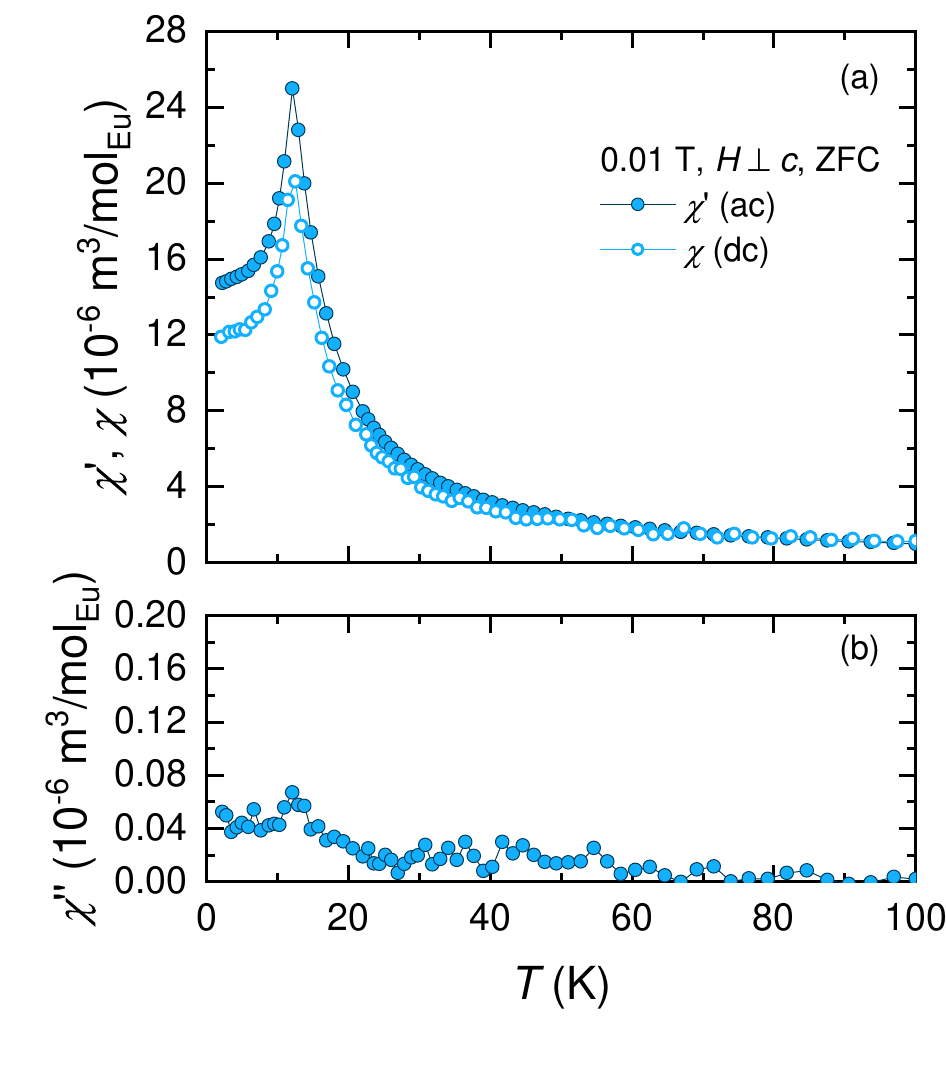}
    \includegraphics[width = 0.49\textwidth]{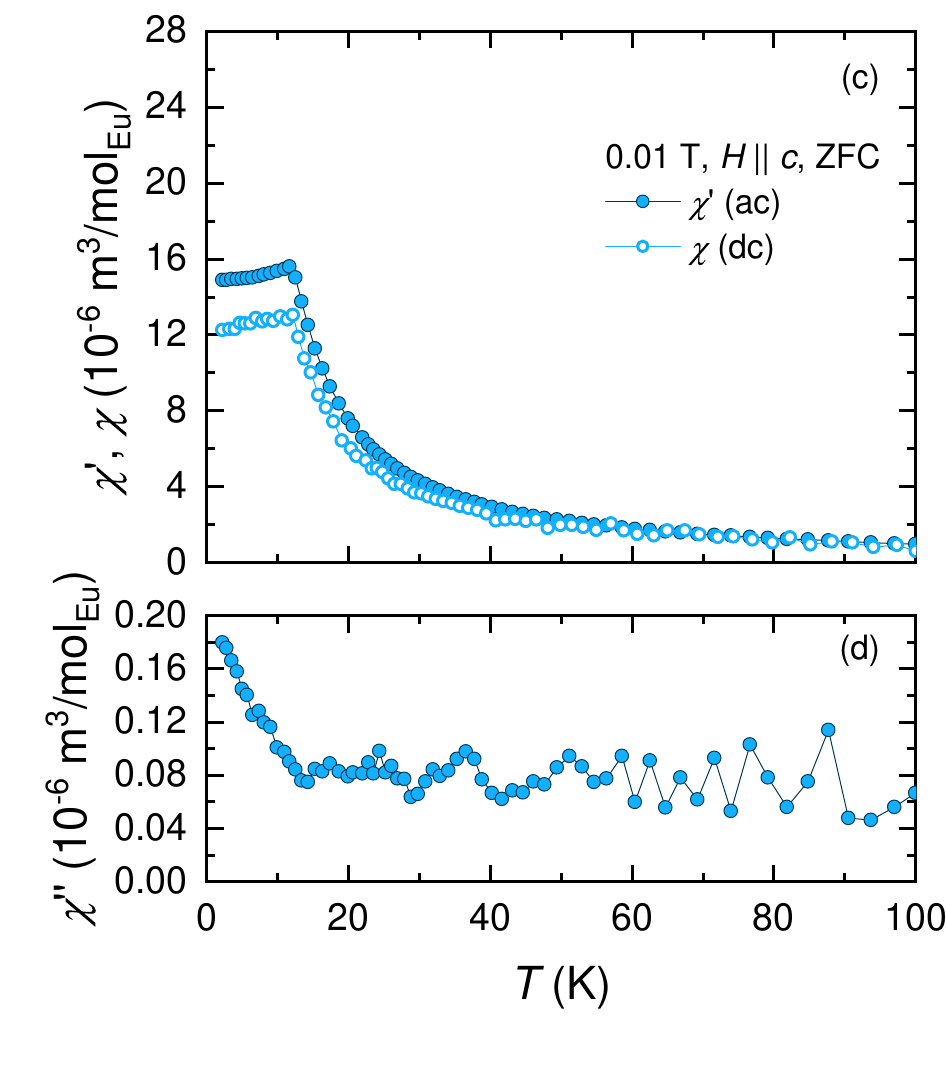}
    \caption{Temperature dependence of ac magnetic susceptibility (solid circles) and dc susceptibility (open circles) measured in $\muH = \SI{0.01}{T}$ magnetic fields applied (a, b) perpendicular and (c, d) parallel to the crystallographic  \cax.}
    \label{fig:ChiT}
\end{figure*}
\begin{figure*}[!ht]
    \centering
    \includegraphics[width=0.49\linewidth]{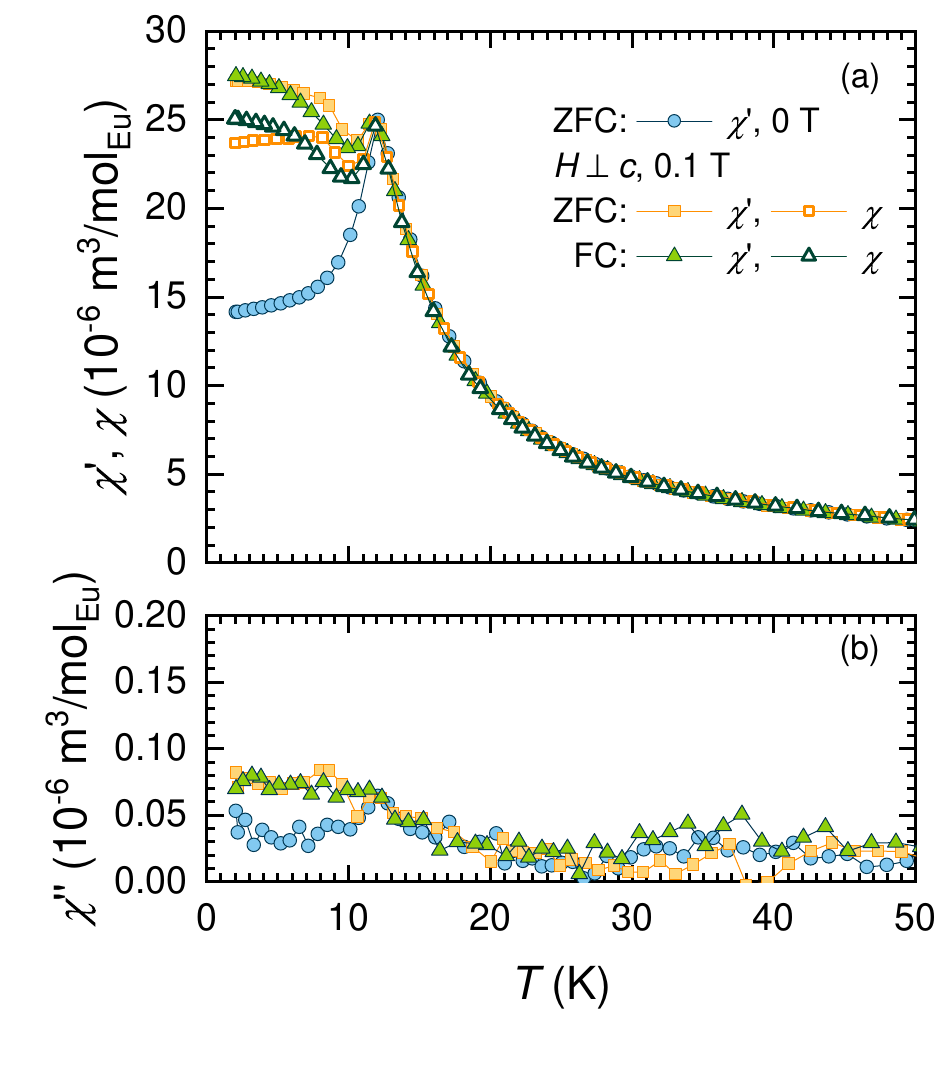}
        \includegraphics[width=0.49\linewidth]{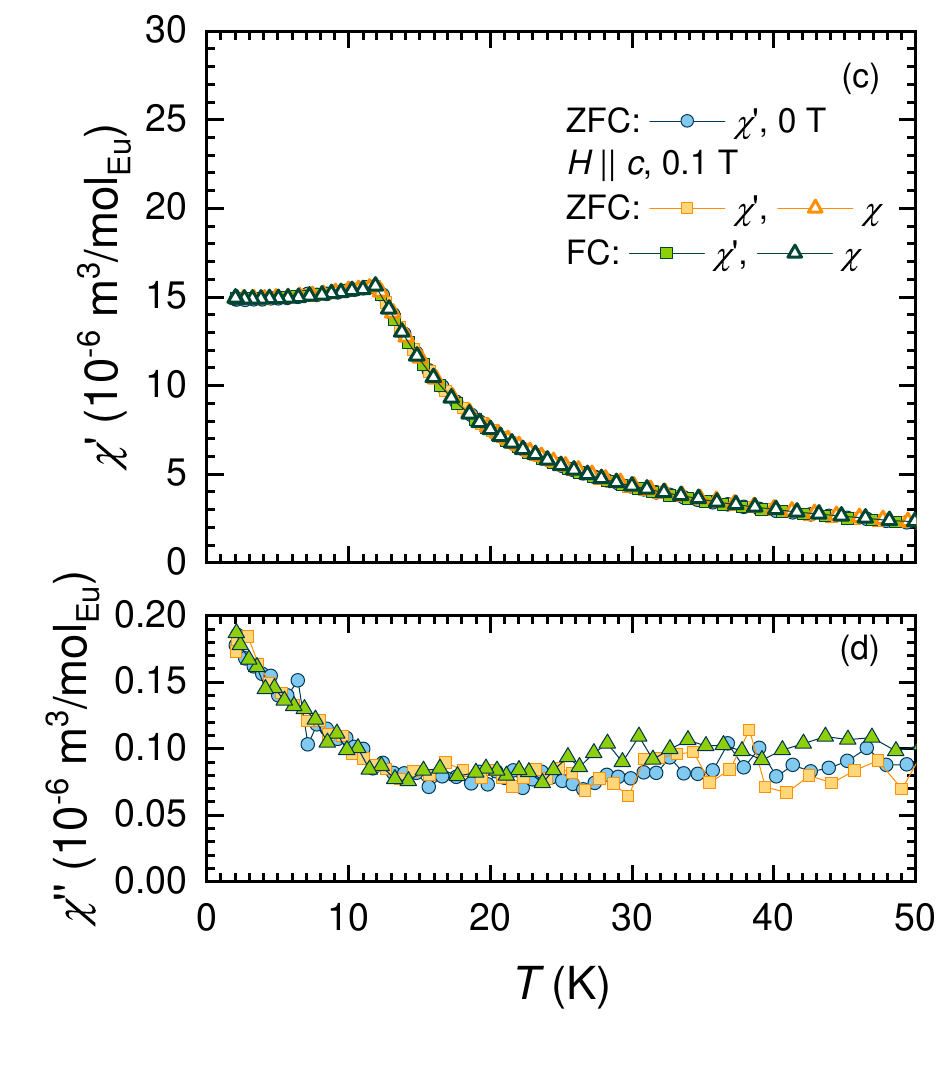}
    \caption{Temperature dependence of ac magnetic susceptibility ($\chi'$, solid symbols)) and dc susceptibility ($\chi$, open symbols) measured in $\muH = \SI{0.1}{T}$ magnetic fields applied (a, b) perpendicular and (c, d) parallel to the crystallographic \cax{} using the ZFC and FC modes.}
    \label{fig:ZFCFC}
\end{figure*}

\begin{figure*}[!ht]
    \centering
    \includegraphics[width = 0.49\textwidth]{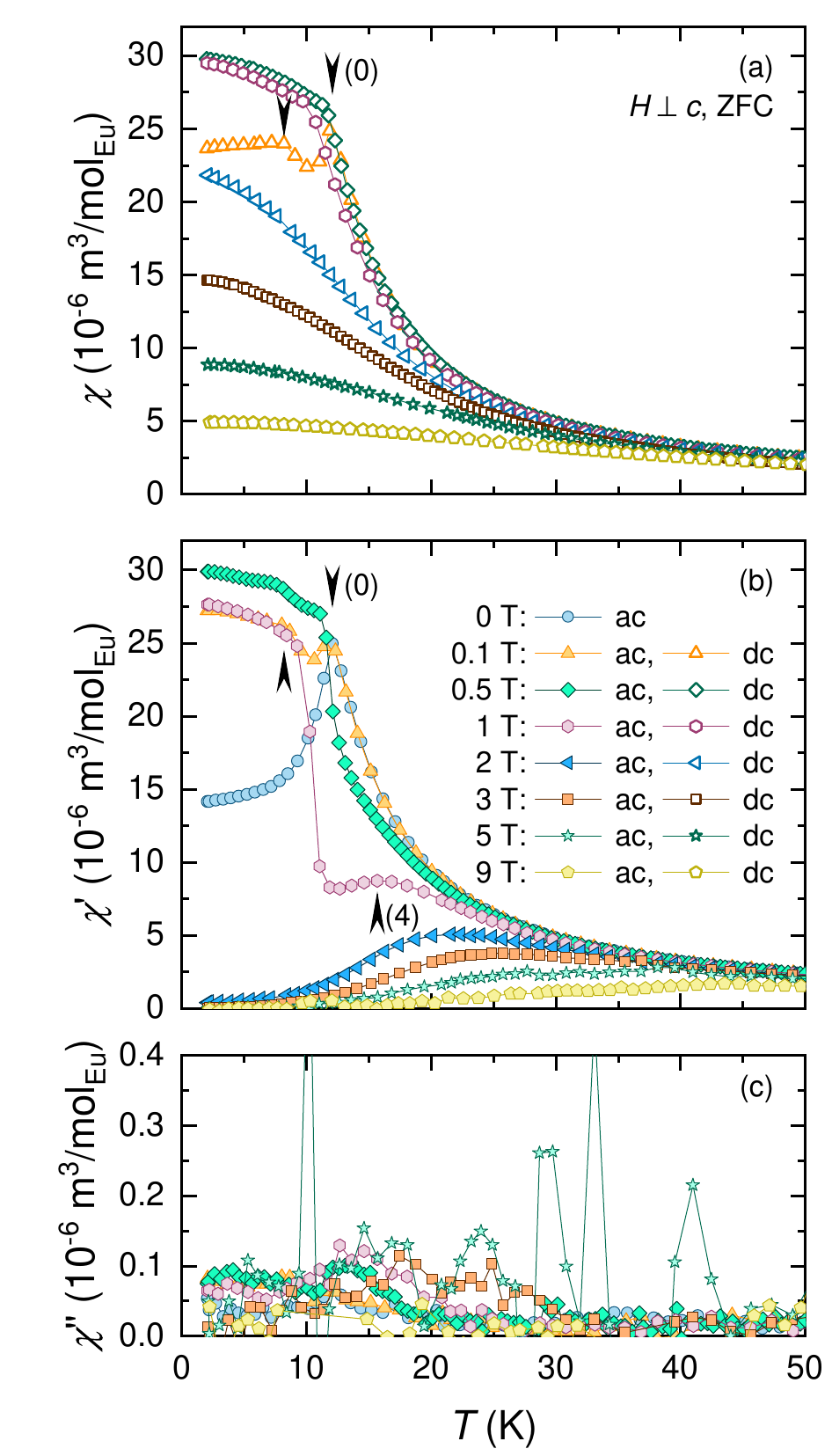}
    \includegraphics[width = 0.49\textwidth]{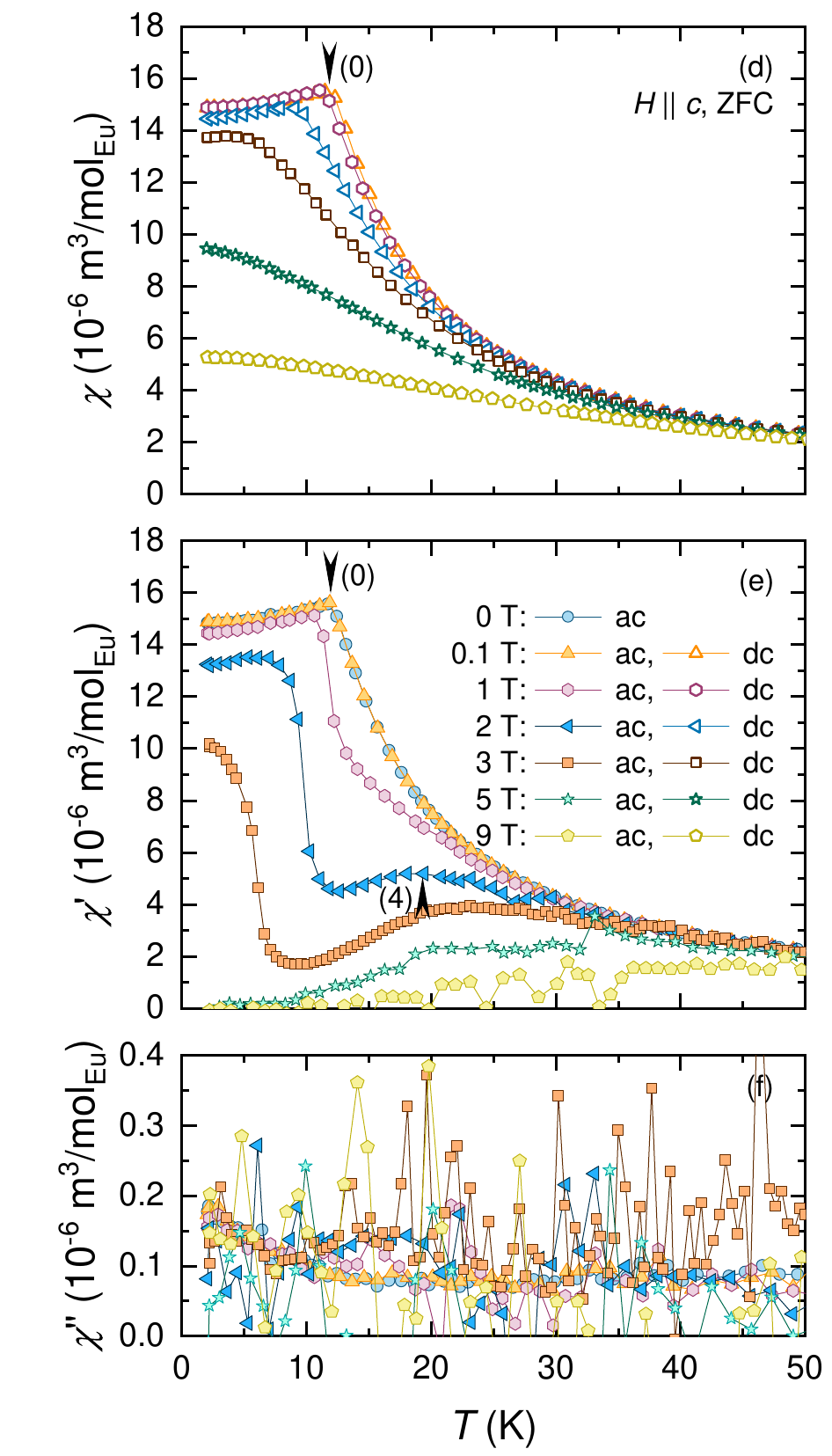}
    \caption{Temperature dependence of dc susceptibility ($\chi$, open symbols) and ac susceptibility (real $\chi'$ and imaginary $\chi''$ parts, solid symbols) measured in several external magnetic fields applied (a, b) perpendicular and (c, d) parallel to the crystallographic $c$ axis. Arrows (0), (4) indicate the characteristic temperatures.}
    \label{fig:Chi-FIF}
\end{figure*}

The temperature dependence of the ac magnetic susceptibility and dc magnetization were investigated using two modes: (ZFC) zero-field-cooled mode, where the sample was cooled down to the lowest temperature in zero magnetic field and then the magnetic field was applied and the measurements were performed while warming the sample; and (FC) field-cooled mode, where the measurements were performed while cooling the sample in the presence of an applied magnetic field.

The antiferromagnetic transition (for external fields $\muH\leq \SI{0.01}{T}$) is clearly visible as a sharp peak at around \SI{12}{K} in the temperature dependence of both ac and dc susceptibility, as shown in Fig.~\ref{fig:ChiT}. This is in agreement with the literature data~\cite{Laha2021} as well as with the heat capacity results discussed in Sec.~\ref{sec:HC}.

For external fields $\muH\leq \SI{2}{T}$ applied perpendicularly to the crystallographic \cax{} (\Hperpc) some splitting between the ZFC and FC branches is observed (see Fig.~\ref{fig:ZFCFC}), suggesting the presence of a ferromagnetic component in the magnetic order of EuAgAs. With the application of magnetic field, the antiferromagnetic peak becomes less sharp and shifts to lower temperatures (a shift expected for an antiferromagnetic transition). Moreover, for $\muH = \SI{0.1}{T}$ below $\TN$ a dip and another (broad) maximum are observed. The intensity of this maximum increases significantly with increasing magnetic field, and as a result obscures the original antiferromagnetic peak (Fig.~\ref{fig:ZFCFC}). 

For magnetic fields applied parallel to the \cax{} (\Hparac), the antiferromagnetic peak also  becomes less sharp and shifts to lower temperatures with increasing magnetic field. However, the splitting between the ZFC and FC curves is not observed (see Fig.~\ref{fig:ZFCFC}), which suggests that a ferromagnetic component is not present for this direction of the external magnetic field.


For fields $\muH \geq \SI{1}{T}$ applied perpendicular or parallel to the \cax{}, another (broad) maximum appears above the antiferromagnetic transition, marked with arrow (4) in Fig.~\ref{fig:Chi-FIF}. This maximum shifts to higher temperatures with increasing magnetic field, which is characteristic of a ferromagnetic contribution to the magnetic response. Such a response can be associated with a crossover between the antiferromagnetic order and a state where the magnetic moments are progressively aligned along the direction of the applied magnetic field, commonly referred to in the literature as a "magnetic-field-induced ferromagnetic state", a "magnetic-field-induced field-polarized state" or a "field-polarized state" (FIF).~\cite{Bauer2017,Bannenberg2018,Bauer2022, Podgorska2025,Rybicki2024,HuaiTran2026}

\begin{figure*}[!ht]
    \centering
    \includegraphics[width=0.49\textwidth]{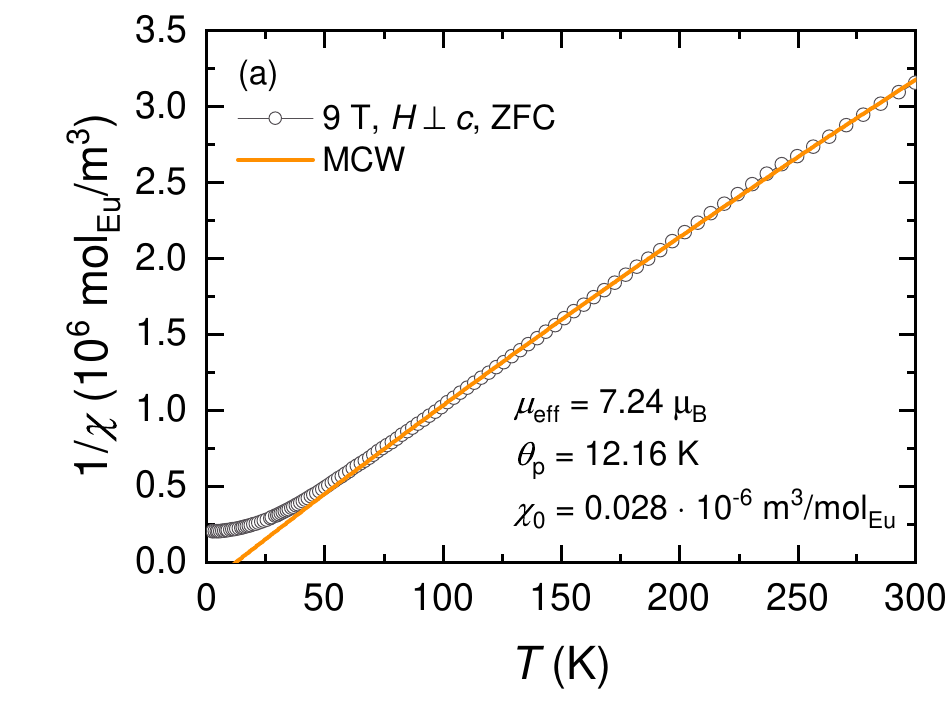}
    \includegraphics[width=0.49\textwidth]{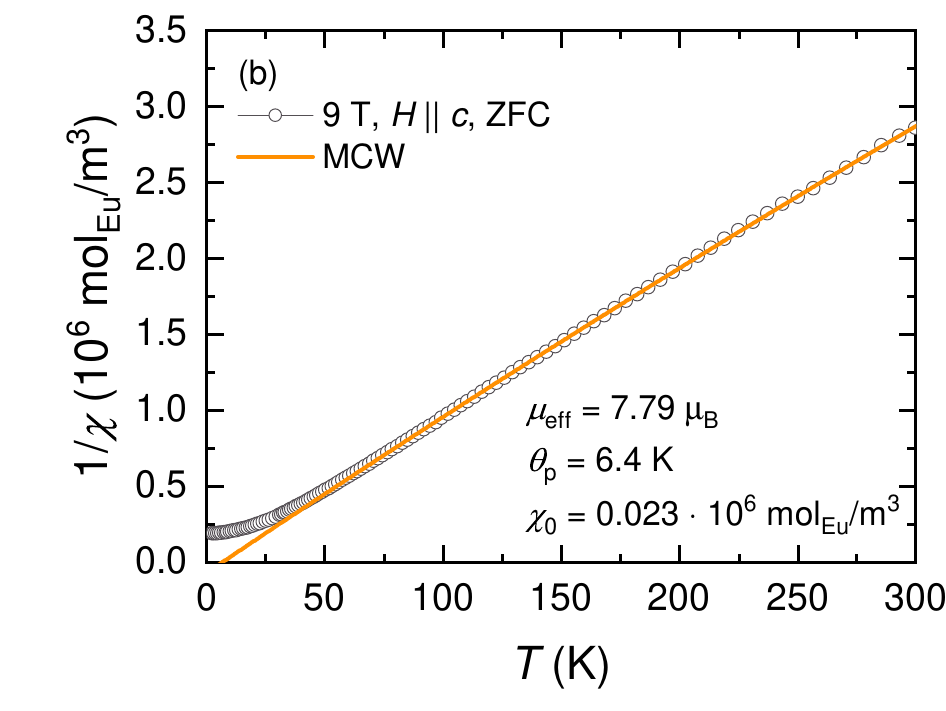}
    \caption{Inverse of the dc magnetic susceptibility, $1/\chi$, as a function of temperature for the \SI{9}{T} external magnetic fields applied (a) perpendicular and (b) parallel to the crystallographic $c$ axis. The solid lines represent the fit of the modified Curie-Weiss law (MCW, cf. Eq. \ref{eq:CW}) to the experimental data in the temperature range from \SI{70}{K} to \SI{300}{K}.}
    \label{fig:CW}
\end{figure*}
The temperature dependence of the magnetic susceptibility in the paramagnetic state can be described by the modified Curie-Weiss law, which is given by the following formula:
\begin{equation}
    \chi(T) = \dfrac{\NA}{3\kB\muO} \dfrac{\mueff^2}{T-\Thp} + \chi_0,
    \label{eq:CW}    
\end{equation}
where $\NA$ is the Avogadro number, $\kB$ is the Boltzmann constant, $\muO$ is the vacuum permeability, $\mueff$ is the effective magnetic moment (in Bohr magnetons \si{\muB}), $\Thp$ is the Weiss temperature (or paramagnetic Curie temperature), and $\chi_0$ is a temperature independent contribution to the magnetic susceptibility. In Fig.~\ref{fig:CW} the inverse of the $\SI{9}{T}$ magnetic susceptibility, $1/\chi$, is plotted as a function of temperature for both directions of the external magnetic field. The solid lines represent the fit of the modified Curie-Weiss law to the experimental data in the temperature range from \SI{70}{K} to \SI{300}{K}, which is well above the antiferromagnetic transition temperature $\TN$. The obtained fitting parameters are summarized in Fig.~\ref{fig:CW}. The calculated effective magnetic moments ($\mueff^{\perp c} = \SI{7.24}{\muB}$, $\mueff^{\parallel c} = \SI{7.79}{\muB}$) are slightly smaller, but close to the theoretical value of \SI{7.94}{\muB} for a free \ce{Eu^{2+}} ion with $S=7/2$ and $L=0$. The positive value of the Weiss temperature ($\Thp^{\perp c} = \SI{12.16}{K}$, $\Thp^{\parallel c} = \SI{6.4}{K}$) indicates that the interactions between the nearest magnetic moments are ferromagnetic, which is consistent with the presence of a ferromagnetic component in the magnetic order of EuAgAs, as suggested by the splitting of the ZFC and FC curves for \Hperpc{}.

\subsubsection{Field dependence}

Field dependences of ac magnetic susceptibility and dc magnetization were investigated at several temperatures, both in external magnetic fields applied parallel and perpendicular to the \cax{} of the crystal. 
For field dependent measurements, the history of the sample was taken into account: (V) first, the initial (virgin) magnetization curve was measured up to \SI{9}{T}, (D) then measurements were carried out with decreasing the magnetic field, (I) and finally with increasing the magnetic field again.

\begin{figure*}[!ht]
    \centering
    \includegraphics[width = 0.49\textwidth]{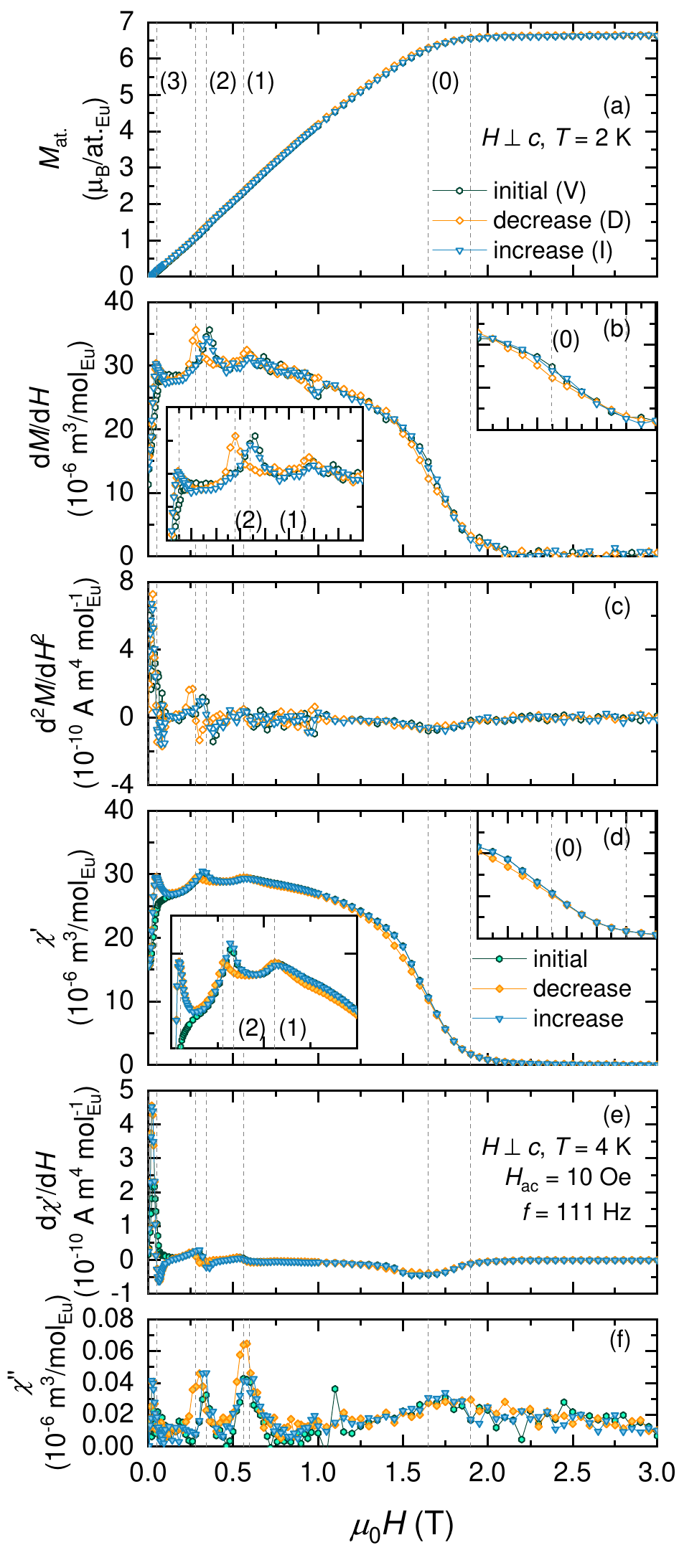}
    \includegraphics[width = 0.49\textwidth]{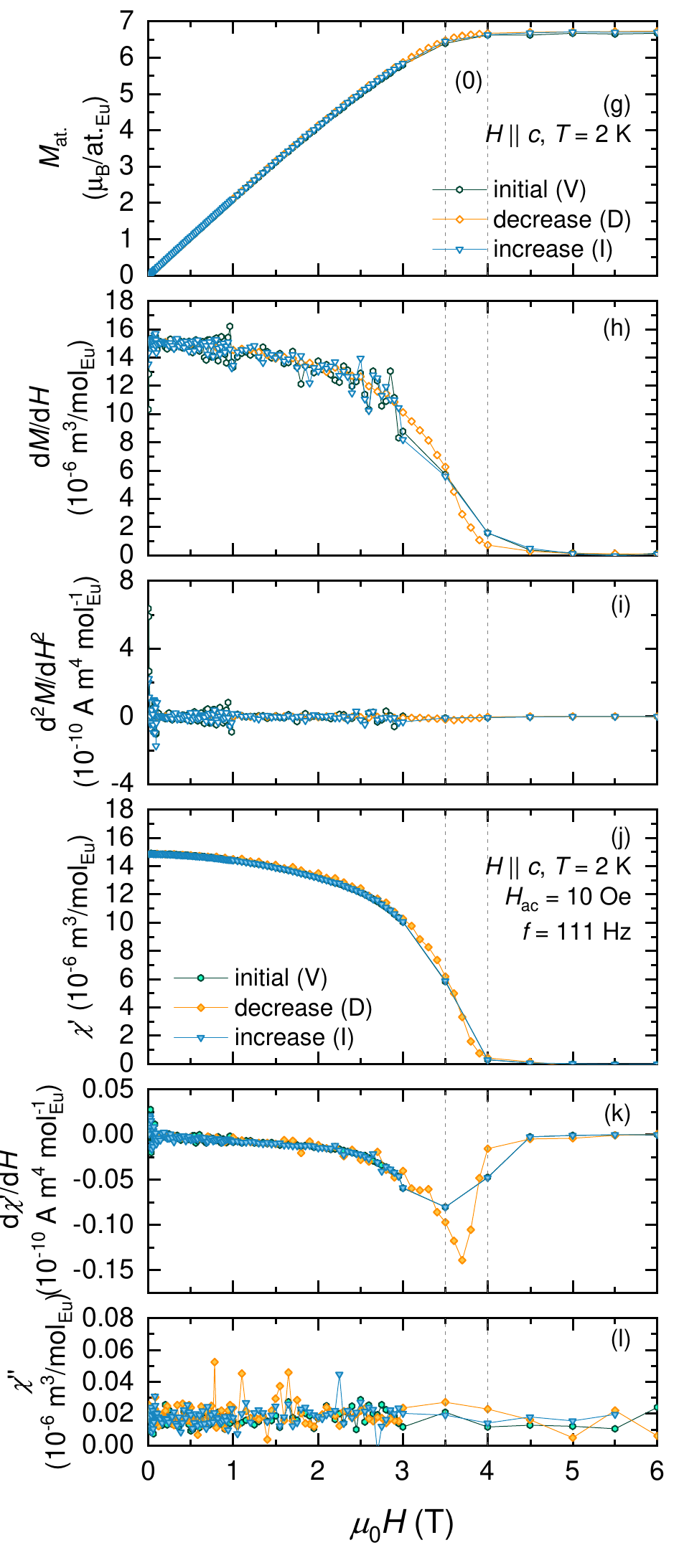}
    \caption{Field dependencies of (a, g) dc magnetization $M\text{at.}(\muH)$, (b, h) dc differential susceptibility $\dd M/\dd H$, (c, i) second derivative of dc magnetization $\dd^2 M/\dd H$, (d, j) real part of ac magnetic susceptibility $\chi'(\muH)$, (e, k) first derivative of the real part of ac susceptibility $\dd \chi'/\dd H$, and (f, l) imaginary part of the ac susceptibility $\chi''$ at \SI{2}{K} in external magnetic fields applied (a-f) perpendicular and (g-l) parallel to the $c$ axis. The numbered (0-3) dashed lines indicate the characteristic fields of the metamagnetic transitions. Insets are zoomed-in views.}
    \label{fig:acdc2K}
\end{figure*}

At a first glance, the \SI{2}{K} field dependence of magnetization $\Mat(\muH)$, Fig.~\ref{fig:acdc2K}(a, g), does not exhibit spontaneous magnetization at zero field, which is consistent with antiferromagnetic order. The initial increase of magnetization with increasing magnetic field appears to be approximately linear. However, after closer inspection, for fields applied perpendicular to the \cax{}, the low-field dependence deviates from linearity, and several metamagnetic transitions accompanied by weak hysteresis can be observed at fields indicated by dashed lines (1)-(3) in Fig.~\ref{fig:acdc2K}(a). This is especially visible in the differential susceptibility $\dd M/\dd H$, shown in Fig.~\ref{fig:acdc2K}(b), where each deviation from the linear field dependence of magnetization can be easily identified as a peak. The hysteretic behavior is particularly evident in the insets of Fig.~\ref{fig:acdc2K}(b), where the field-increasing and field-decreasing branches can be clearly distinguished. Similar peaks (and hysteretic behavior), while not as prominent, are clearly visible in the real part of the ac susceptibility $\chi'(\muH)$, shown in Fig.~\ref{fig:acdc2K}(d). This behavior resembles that observed in some materials with a helical magnetic structure, where the application of an external magnetic field can induce a transition from a helical to a conical state, which is often accompanied by a peak in the ac (and differential dc) susceptibility as a function of field and a change in the slope of the magnetization curve~\cite{Bannenberg2018, Bauer2022}.
The difference in intensity  between $\dd M/\dd H(\muH)$ and $\chi'(\muH)$ suggests that the reorientation occurs over a macroscopic time scale~\cite{ToppingBlundell2019, Bauer2017, Bannenberg2018, Bauer2022}. 

In comparison, the values of $\dd M/\dd H$ for fields applied parallel to the \cax{} are smaller, indicating smaller deviations from the linear dependence of $\Mat(\muH)$, and no apparent peaks are visible. Similarly, no anomalies are observed in the ac susceptibility $\chi'(\muH)$. 

At about \SI{1.6}{T} for fields applied perpendicular to the \cax{} and at about \SI{3.5}{T} for fields applied parallel to the \cax{}, indicated by the dashed line labeled (0), a crossover associated with the reorientation of magnetic moments toward the FIF state is observed. Above this metamagnetic transition, magnetization saturates to values of \SI{6.65}{\muB} for \Hperpc{} and \SI{6.75}{\muB} for \Hparac (for $\muH = \SI{9}{T}$), which are close to the expected theoretical value of \SI{7}{\muB} for free \ce{Eu^{2+}} ions. One should note that the saturation is not perfect, suggesting that complete magnetic moments alignment is not achieved at \SI{9}{T}. There can be several possible reasons for this. For example, the applied field may be insufficient to remove all domain boundaries and establish a single-domain state.
%


\subsection{Imaginary part of ac susceptibility}
Although some peaks in the imaginary part of the ac susceptibility $\chi''$ are observed at temperatures (or fields) corresponding to the peaks in the real part $\chi'$, the intensity of these peaks is significantly smaller (by approximately two orders of magnitude). Therefore, it cannot be excluded that the observed peaks in the imaginary part are due to an experimental artifact, such as a tiny, well below \SI{1}{\degree}, phase shift between the driving field and the response of the sample, 
which can lead to a small contribution of the real part of the susceptibility to the imaginary part.
Nevertheless, the small imaginary part of the ac susceptibility indicates that the magnetic reorientation process is associated with weak dissipative losses.

\section{Summary and conclusions}

\subsection{Magnetic phase diagram}\label{sec:PhDiag}

\begin{figure*}[!hbt]
    \centering
    \includegraphics[width=1\linewidth]{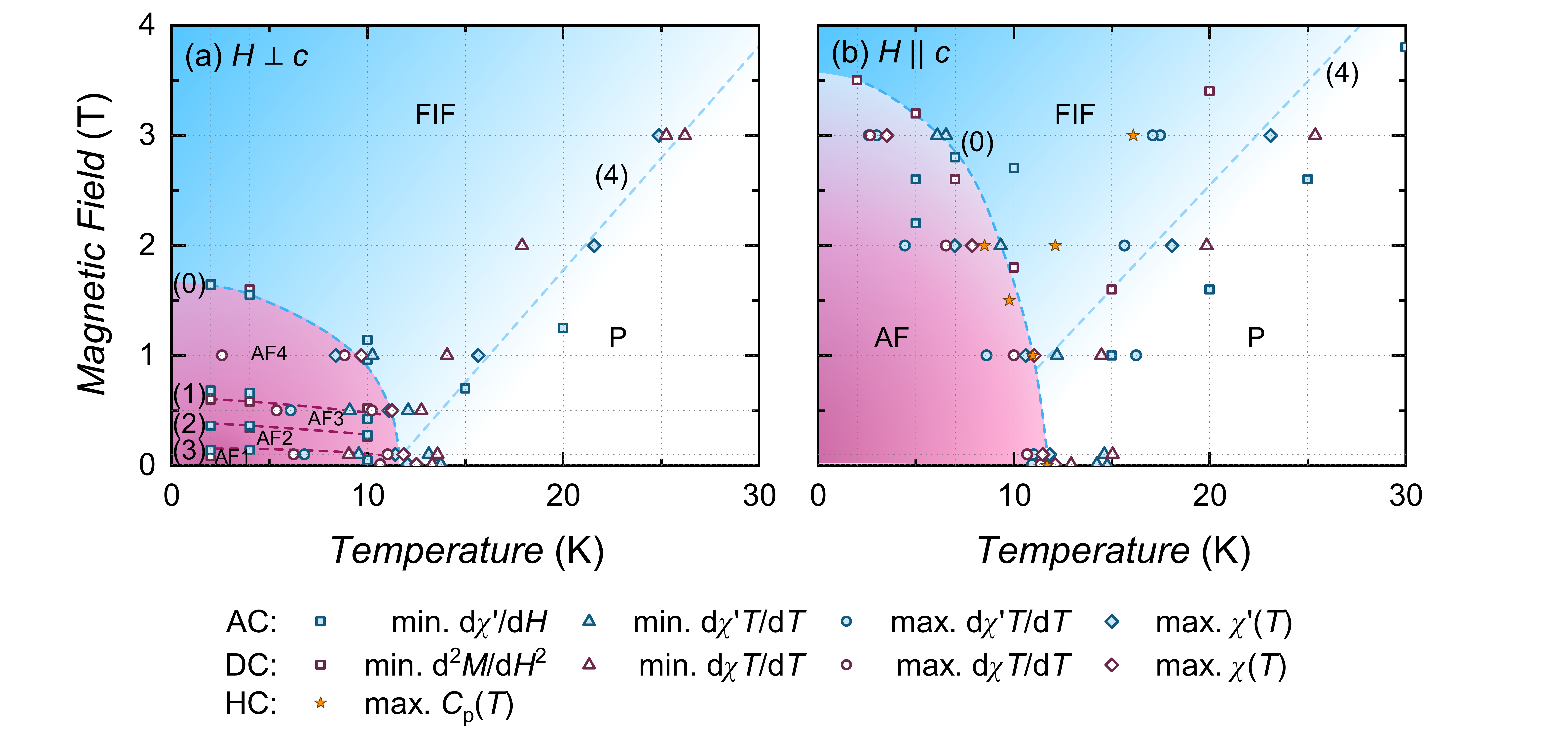}
    \caption{Magnetic phase diagram of \ce{EuAgAs} constructed based on magnetic and heat capacity measurements. 
    The gray dots indicate the points at which the signal was recorded.
    The dashed lines are guides for the eye.
    The numbers (0-4) indicate the characteristic temperatures and fields, which were determined from the ac susceptibility (AC, solid points), dc magnetization (DC, open points) and heat capacity (HC, stars) as the positions of minima (min.) and maxima (max.) in the measured quantities or their derivatives, as indicated in the legend.
    }
    \label{fig:PhDiag}
\end{figure*}

Based on the results of the magnetic and heat capacity measurements, we constructed the magnetic phase diagram of EuAgAs, which is shown in Fig.~\ref{fig:PhDiag}. The points in the phase diagram were determined from the characteristic features observed in the temperature and field dependencies of the ac susceptibility, dc magnetization, and heat capacity. The characteristic points were identified as the positions of minima and maxima in the measured quantities or their derivatives, as indicated in the legend of Fig.~\ref{fig:PhDiag}. Several transitions, labeled (0)-(4), can be distinguished. The transitions (0) and (4) are visible for both directions of the applied magnetic field. The transition (0) corresponds to the boundary between the antiferromagnetic (AF) state and either the paramagnetic (P) or "field induced ferromagnetic" (FIF) state, while transition (4) represents a crossover between the P and FIF states. The transitions (1)-(3) correspond to metamagnetic transitions observed at low temperatures and lower magnetic fields when the field is applied perpendicular to the \cax{}. These transitions are most likely associated with successive reorientations of the magnetic moments.

EuAgAs is antiferromagnetic (AF, AF1) in its ground state. The overall behavior observed in the magnetic measurements and summarized in the phase diagram resembles that reported for some skyrmion hosting materials, in which several metamagnetic transitions occur at low temperatures and low magnetic fields and are associated with the formation of distinct magnetic phases, such as helical, conical, and skyrmion lattice phases~\cite{Bannenberg2018, Bauer2022, HuaiTran2026}. Therefore, based on these observations (e.g., splitting of the ZFC and FC branches, differences in the intensities of the peaks in the field dependence of the differential susceptibility and ac susceptibility, and the tilting of the magnetic moments observed in M\"ossbauer spectroscopy), we speculate that \ce{EuAgAs} in the AF and AF1 regions has a non-collinear antiferromagnetic (possibly helical) structure. With the application of an external magnetic field perpendicular to the \cax{}, in region AF2 or AF3, \ce{EuAgAs} may undergo a transition to a skyrmionic state (or another complex magnetic state), followed by a transition to another non-collinear antiferromagnetic state (AF3 and/or AF4). Given the metallic character of EuAgAs, the magnetic order is expected to be governed predominantly by Ruderman-Kittel-Kasuya-Yosida (RKKY) interactions. Therefore, the possibility of a skyrmionic phase is particularly intriguing in the context of centrosymmetric rare-earth intermetallic compounds, where competing RKKY interactions can stabilize non-collinear and topologically non-trivial spin textures, including skyrmion lattices~\cite{HuaiTran2026}. 





\subsection{Conclusions}
In summary, we have systematically investigated the structural, magnetic, and local electronic properties of \ce{EuAgAs} using x-ray diffraction, M\"{o}ssbauer spectroscopy, dc magnetization, ac susceptibility, and heat capacity measurements. Temperature-dependent x-ray diffraction measurements confirmed that \ce{EuAgAs} crystallizes in the hexagonal $P6_3/mmc$ structure and does not undergo any structural phase transition between 15 and \SI{300}{K}. The lattice parameters decrease monotonically upon cooling, and the unit-cell volume follows Debye-type thermal contraction behavior, indicating conventional lattice dynamics without detectable structural anomalies in the investigated temperature range. 
The M\"{o}ssbauer spectroscopy measurements indicate that Eu electronic state is as for \Euion{} ions. The magnetic and thermodynamic measurements confirm antiferromagnetic ordering below $\TN \approx \SI{12}{K}$ and reveal pronounced magnetic anisotropy, as well as several field-induced metamagnetic transitions for magnetic fields applied perpendicular to the \cax{}. Based on the characteristic features observed in the magnetic and heat-capacity measurements, we constructed the magnetic phase diagram of \ce{EuAgAs} and identified several distinct magnetic regions.
The splitting of the ZFC and FC branches, together with the anomalies observed in the field dependence of the differential susceptibility and ac susceptibility, suggests that the low-field antiferromagnetic state may possess a noncollinear magnetic structure, possibly of helical character. The sequence of field-induced transitions observed for fields perpendicular to the \cax{} resembles the behavior reported for several centrosymmetric RKKY-interaction-driven rare-earth  compounds in which competing magnetic interactions can stabilize noncollinear and skyrmion phases. Although the present measurements do not provide direct evidence for skyrmion formation in EuAgAs, they point to a potentially complex field-induced magnetic state. Furthermore, the M\"{o}ssbauer spectroscopy results reveal that the \Euion{} magnetic moments are tilted with respect to the crystallographic \cax{}, supporting a departure from a simple collinear magnetic structure. Our results provide a comprehensive characterization of the complex magnetic phase diagram of \ce{EuAgAs} and suggest a noncollinear magnetic ground state. Further microscopic investigations, particularly neutron diffraction or Lorentz transmission electron microscopy, would be valuable for determining the magnetic structures associated with the different regions of the phase diagram.




\section*{Data availability}
The datasets used and/or analyzed during the current study are available from the corresponding author on reasonable request.

\section*{Acknowledgments}

We acknowledge financial support by National Science Centre, Poland (Grants No. 2018/30/E/ST3/00377 and 2021/41/B/ST3/03454). The research project was partly supported by the program “Excellence initiative–research university” for the AGH University of Krakow and by a subsidy from the Polish Ministry of Science and Higher Education.

\section*{Author Contributions}
KP: 
conceptualization, 
synthesis of single crystals, 
investigation (participation in all measurements), formal analysis, 
visualization of data, 
writing-original draft. 
KK: 
formal analysis, 
investigation (M\"ossbauer spectroscopy measurements). 
JP: investigation (heat capacity and VSM measurements), 
formal analysis. 
ŁG: 
investigation (XRD and SEM measurements), 
formal analysis, 
visualization of data.  
CK: 
writing-review. 
WT: 
supervision and review. 
MB: 
conceptualization, 
synthesis of single crystals, 
investigation (magnetic properties measurements), 
formal analysis, 
writing-original draft, 
supervision. 
LMT: 
conceptualization,
formal analysis, 
visualization of data, 
writing (original draft, review \& editing), 
investigation (magnetic properties measurements). 
DR: 
conceptualization, 
investigation (M\"ossbauer spectroscopy measurements), supervision, 
writing-original draft. 
All authors contributed to the article and approved the submitted version.
\bibliography{sample}

\end{document}